\documentclass[12pt]{article}

\usepackage{amsmath}
\usepackage{amsfonts}
\usepackage{amssymb}
\usepackage{graphics,graphicx,tikz}
\usepackage{subfig}
\usepackage{graphicx}
\usepackage{soul}
\usepackage{cite}
\usepackage{graphicx}
\usetikzlibrary{patterns}
\usepackage{enumitem}
\usepackage{array}
\usepackage{amsbsy}
\usepackage{bbold}
\usepackage{appendix}
\usepackage{mathrsfs}
\usepackage{physics}
\usepackage{amsthm}
\usepackage{cancel}
\usepackage{yfonts}
\usepackage{inputenc}
\usepackage{chngcntr}

\numberwithin{equation}{section} 
\usepackage[linktocpage]{hyperref}
\hypersetup{
	colorlinks=true,
	linkcolor=blue,
	filecolor=magenta,      
	urlcolor=blue,
	citecolor=blue
}
\def\beq{\begin{equation}}
\def\eeq{\end{equation}}
\newcommand{\commentOut}[1]{}
\def\bea{\begin{align}}
\def\eea{\end{align}}

\usepackage{color}

\begin{document}
\begin{titlepage}
\vskip 0.1cm
\vskip 1.5cm
\begin{flushright}
\end{flushright}
\vskip 1.0cm
\begin{center}
{\Large \bf 2PM waveform from loop corrected soft theorems}

\vskip 1.0cm {\large  Francesco Alessio$^{a,b}$  and Paolo Di Vecchia$^{a, c}$  } \\[0.7cm]

{\it \small $^a$ NORDITA, KTH Royal Institute of Technology and Stockholm University, \\
 Hannes Alfv{\'{e}}ns v{\"{a}}g 12, SE-11419 Stockholm, Sweden  }\\
 {\it \small $^b$ Department of Physics and Astronomy, Uppsala University,\\ Box 516, SE-75120 Uppsala, Sweden}

{\it \small $^c$ The Niels Bohr Institute, Blegdamsvej 17, DK-2100 Copenhagen, Denmark}\\

\end{center}
\begin{abstract}
We introduce a classical version of the loop corrected soft graviton theorem and we use it to compute the universal part of the one-loop (2PM) waveform up to sub-subleading order in the energy $\omega$ of the emitted graviton for spinless black holes scattering. In particular, we compute the action of the soft operators on the classically resummed four-point amplitude, that can be written in terms of the exponential of the eikonal phase (and is therefore non-perturbative in the Newton's constant $G_N$) and then we perform the usual PM expansion in powers of $G_N$. We find perfect agreement with the existing 2PM literature at the orders $\omega^{-1}$, $\log\omega$ and $\omega \log^2\omega$, which are universal. Furthermore, we use this method to compute the universal part of the $\omega\log\omega$ contribution to the 2PM waveform. Even if in the present analysis we limit ourselves to compute the soft 2PM waveform, our general formulae can be used to extract all universal PM orders of the terms connected with  the infrared divergences, once the impulse at the corresponding precision is known. Our approach, based on the resummed eikonal amplitude, gives a unified picture of the various computations of the classical soft graviton behaviour that are present in the literature since the seminal paper by Weinberg in 1965 \cite{Weinberg:1965nx}.
\end{abstract}
\end{titlepage}


\section{Introduction}
\label{intro}

In the last few years  techniques based on scattering amplitudes have been successfully applied to the computation of gravitational observables  in the framework of 
the  Post-Minkowskian (PM) expansion. They  have been used to extract classical observables from the full quantum amplitude. A non-exhaustive list of PM literature is \cite{Buonanno:1998gg,Goldberger:2004jt,Goldberger:2016iau,Cachazo:2017jef,Luna:2017dtq,Bjerrum-Bohr:2018xdl,Cheung:2018wkq,Kosower:2018adc,Bern:2019nnu,Bern:2019crd,Brandhuber:2019qpg,AccettulliHuber:2019jqo,KoemansCollado:2019ggb,Cristofoli:2019neg,Bjerrum-Bohr:2019kec,Cheung:2020gyp,Parra-Martinez:2020dzs,Brandhuber:2021eyq,Bern:2021dqo,Cristofoli:2020uzm,AccettulliHuber:2020dal,delaCruz:2020bbn,Cristofoli:2021vyo,Herrmann:2021tct,Cristofoli:2021jas,Damour:2017zjx,Herrmann:2021lqe,DiVecchia:2020ymx,DiVecchia:2021ndb,DiVecchia:2021bdo,DiVecchia:2022nna,DiVecchia:2022piu,Bjerrum-Bohr:2021vuf,Bjerrum-Bohr:2021din,Bjerrum-Bohr:2021wwt,Bern:2021yeh,Kalin:2020mvi,Kalin:2020fhe,Dlapa:2021npj,Dlapa:2021vgp,Kalin:2022hph,Dlapa:2022lmu,Mogull:2020sak,Jakobsen:2022psy,Khalil:2022ylj,Jones:2022aji,Bini:2022wrq,Brandhuber:2023hhy,Georgoudis:2023lgf,Arkani-Hamed:2017jhn,Guevara:2017csg,Bini:2017xzy,Vines:2017hyw,Bini:2018ywr,Vines:2018gqi,Guevara:2018wpp,Chung:2018kqs,Bautista:2019tdr,Bautista:2019evw,Maybee:2019jus,Guevara:2019fsj,Arkani-Hamed:2019ymq,Johansson:2019dnu,Chung:2019duq,Damgaard:2019lfh,Chung:2019yfs,Aoude:2020onz,Chung:2020rrz,Bern:2020buy,Aoude:2020ygw,Guevara:2020xjx,Liu:2021zxr,Kosmopoulos:2021zoq,Aoude:2021oqj,Jakobsen:2021lvp,Bautista:2021wfy,Chiodaroli:2021eug,Haddad:2021znf,Gonzo:2021drq,Jakobsen:2021zvh,Saketh:2021sri,Adamo:2021rfq,Chen:2021qkk,Jakobsen:2022fcj,Aoude:2022trd,Aoude:2022thd,Bern:2022kto,Alessio:2022kwv,Chen:2022clh,Ochirov:2022nqz,Damgaard:2022jem,FebresCordero:2022jts,Menezes:2022tcs,Riva:2022fru,Cangemi:2022abk,Hung:2022azf,Jakobsen:2022zsx,Saketh:2022wap,Bjerrum-Bohr:2023jau,Bautista:2022wjf,Cangemi:2022bew,Comberiati:2022ldk,Comberiati:2022cpm,Kim:2023drc,A:2022wsk,Hoogeveen:2023bqa,Haddad:2023ylx,Elkhidir:2023dco,Bhattacharyya:2024aeq}. These include matching QFT results with an EFT description~\cite{Neill:2013wsa,Bjerrum-Bohr:2018xdl,Cheung:2018wkq}, developing the worldline formalism~\cite{Kalin:2020mvi,Mogull:2020sak} to compute higher orders in the PM expansion  and using the KMOC approach~\cite{Kosower:2018adc} or the eikonal exponentiation~\cite{Amati:1987wq,Amati:1987uf,KoemansCollado:2019ggb,DiVecchia:2023frv} for computing observables for both spinless and spinning particles. The scattering results have then also been used in EOB models~\cite{Buonanno:1998gg,Buonanno:2000ef,Khalil:2022ylj,Damour:2022ybd,Rettegno:2023ghr} and for bound orbits  \cite{Kalin:2019rwq,Kalin:2019inp,Cho:2021arx,Gonzo:2023goe,Adamo:2024oxy}.

A major breakthrough was the two-loop calculation~\cite{Bern:2019nnu,Bern:2019crd} of the impulse in the potential region  completed  with  the calculation~\cite{DiVecchia:2020ymx,Damour:2020tta,DiVecchia:2021ndb,DiVecchia:2021bdo,Alessio:2022kwv} of the radiation reaction that made the deflection angle  finite at high energy. 
Those results were confirmed by a direct calculation~\cite{Bjerrum-Bohr:2021vuf,Bjerrum-Bohr:2021din,Herrmann:2021tct,Brandhuber:2021eyq} of the two-loop  classical elastic amplitude.
During the last couple of years three groups~\cite{Dlapa:2021vgp,Dlapa:2022lmu,Damgaard:2023ttc,Jakobsen:2023ndj,Jakobsen:2023hig} have computed both the contribution of the potential region and that of the radiation reaction to the deflection angle at 4PM in the elastic scattering of  two massive particles obtaining identical results.  In particular, Refs.~\cite{Jakobsen:2023ndj,Jakobsen:2023hig}  included also the contribution of few powers of the spin.

One of the main objectives of theoretical work on gravitational waves is to compute the waveform of the gravitational waves emitted by the merging or by  the scattering of two black holes. The leading waveform was already computed in a pioneering work in Ref.~\cite{Kovacs:1978eu} using standard general relativity methods. Recently it has been computed using 
scattering amplitude techniques both in time domain~\cite{Jakobsen:2021smu} and in frequency domain~\cite{Mougiakakos:2021ckm,Mougiakakos:2022sic,Riva:2022fru}. Generalization to the spinning case can be found in \cite{DeAngelis:2023lvf,Aoude:2023dui,Brandhuber:2023hhl}. Those papers provide
the waveform at 1PM order, that is obtained from the classical tree-level $5$-point amplitude involving four massive particles and a graviton.  Recently, the $5$-point amplitude and the waveform at 2PM (one-loop) have also been computed~\cite{Brandhuber:2023hhy,Herderschee:2023fxh,Georgoudis:2023lgf,Elkhidir:2023dco} (see also  the subsequent 
papers~\cite{Caron-Huot:2023vxl,Bini:2023fiz,Georgoudis:2023eke,Bohnenblust:2023qmy,Georgoudis:2023ozp}).  In particular, Refs.~\cite{Bini:2023fiz,Georgoudis:2023eke} deal with the one-loop waveform in the limit in which the graviton is soft.

At the tree-level the amplitude with  emission of a soft graviton satisfies low-energy theorems that follow directly from gauge invariance~\cite{Bern:2014vva,Broedel:2014fsa}.
According to these theorems, the leading $\mathcal{O}(\omega^{-1})$ and the subleading $\mathcal{O}(\omega^0)$ terms are universal and
they apply to any theory. The sub-sub-leading $\mathcal{O}(\omega)$ term instead contains a universal piece, but non-universal pieces 
can appear in different cases. 

At one-loop level scattering amplitudes are infrared divergent in spacetime dimensions $D=4$  and they  must be 
regularised by  introducing, for instance,  an infrared regulator $\epsilon= \frac{4-D}{2}$.  Then  the previous low-energy theorems are modified and
contain divergent terms in the $\epsilon \rightarrow 0$ limit ~\cite{Bern:2014oka}. 

Recently in a series of papers~\cite{Laddha:2018vbn,Sahoo:2018lxl,Saha:2019tub,Sahoo:2021ctw,Ghosh:2021bam,Krishna:2023fxg} the authors have considered the  low-energy theorems, computing the new 
universal terms involving $\log \omega$ up to one-loop level. The $\log \omega$ term has also been independently obtained in Refs.~\cite{Addazi:2019mjh,Ciafaloni:2018uwe} using methods based on the eikonal resummation and it has been shown  in Ref.~\cite{Agrawal:2023zea} to follow 
from superrotation Ward identities. It has also been confirmed~\cite{Georgoudis:2023eke} from the soft expansion of the complete one-loop $5$-point amplitude involving four massive particles 
and a graviton. This last paper confirms the universality of the $\mathcal{O}(\omega^{-1})$, $\mathcal{O}(\log\omega)$ and $\mathcal{O}(\omega \log^2 \omega)$ terms, but also shows that the order $\mathcal{O}( \omega\log \omega)$ is not universal, as pointed out \textit{e.g.} in Ref. \cite{Krishna:2023fxg}.   

As shown in both  Refs.~\cite{Bern:2014oka} and \cite{Sahoo:2018lxl,Krishna:2023fxg}, the soft factors involve the calculation of two types of one-loop Feynman diagrams, one where the internal graviton is attached to two massive legs and another  where the internal graviton is attached to a massive and a massless leg. They both need to be regularised. In Ref.~\cite{Bern:2014oka} it is used dimensional regularisation. On the other hand, as done in \cite{Weinberg:1965nx}, one can also regularise them by the introduction of an ultraviolet $\Lambda$ and an infrared cutoff $\lambda$ and then relate the two regularisations by imposing   that $\frac{1}{2\epsilon} = \log \frac{\lambda}{\Lambda}$.  If we then assume that $\omega$ acts as an infrared cutoff
in  the first type of diagram and instead acts as an ultraviolet cutoff in the second type of diagram, we get exactly the $\log \omega$ soft factors obtained in~\cite{Sahoo:2018lxl,Krishna:2023fxg}, as pointed out in~\cite{Agrawal:2023zea}.     

In this paper we follow the procedure outlined in Ref.~\cite{Krishna:2023fxg} (with the limitations discussed there for the $\mathcal{O}(\omega \log \omega)$ term) for computing the various soft factors in the classical limit and then we use them in the case of a  $5$-point amplitude involving four scalar particles and a graviton. However, instead of applying  them to a quantum $4$-point amplitude, we act with them on a resummed classical $4$-point amplitude  that is obtained by transforming back to  momentum space  the eikonal in impact parameter space.  In this way we compute the  $\mathcal{O}(\log \omega)$ and $\mathcal{O}(\omega\log \omega)$ terms of the 1PM waveform and the same ones at 2PM, together with $\mathcal{O}(\omega\log^2\omega)$.

 The paper is organised as follows.  In section \ref{soft} we summarise how the tree-level and one-loop soft theorems are obtained, mainly following the procedure outlined in Refs.~\cite{Sahoo:2018lxl,Krishna:2023fxg}.  In section \ref{eikonal} we revisit the eikonal exponentiation and, in particular, we discuss how the resummed elastic four-point amplitude in momentum space can be written in terms of the eikonal phase. In section \ref{resum} we perform the soft expansion of the tree-level waveform, starting from the soft expansion of the tree-level five-point amplitude. In particular, we compute its $\mathcal{O}(\omega^{-1})$, $\mathcal{O}(\log\omega)$ and $\mathcal{O}(\omega\log\omega)$ contributions. In section \ref{eiksoft} we introduce a soft graviton theorem that acts on the resummed $4$-point amplitude, obtaining the  $5$-point amplitude in the soft limit and up to order 2PM. In section \ref{universal} we write the soft contribution to the waveform both at 1PM and at 2PM orders and in section \ref{PN} we displat the first orders of their PN expansion. Section~\ref{conclu} is devoted to some conclusions. The details of the calculations of the results presented in section \ref{universal} are given in appendix~\ref{A}.    
Finally in Appendix \ref{C} we show how to extract the classical contribution of the Weinberg function $\mathcal{W}$ from the full quantum calculation.   
\subsection*{Note added}
After the submission of the present paper to the arXiv, we became aware of two related papers \cite{Georgoudis:2024pdz,Bini:2024rsy} submitted on the same day with partial overlap with our analysis.

\section{Soft theorems}
\label{soft}
Let us start by considering an elastic scattering process $n\rightarrow m$ involving $n$ incoming and $m$ outgoing massive particles and let us denote the corresponding tree-level scattering amplitude by $\mathcal{M}_{N}$, with $N=n+m$. We introduce the shorthand notation $p_i\equiv (p_1,\dots,p_n)$ and $p_f\equiv (p_{n+1},\dots,p_m)$ for the incoming and outgoing asymptotic momenta and
\begin{align}
\label{F0}
\eta_j=\left\{\begin{matrix}-1\hspace{1cm} \mathrm{incoming}\\1\hspace{0.7cm}\mathrm{outgoing}\end{matrix}\right.,\qquad \sigma_{jk}=-\frac{p_j\cdot p_k}{m_j m_k},\qquad j,k=1,\dots N.
\end{align}
In our convention, all momenta are outgoing and hence we parametrize them as
\begin{align}
\label{F01}
p_j=\eta_j(E_j,\vec{p}_j),\qquad E_j>0,\qquad p^2_j=-m^2_j.
\end{align}
We are interested in the tree-level amplitude $\mathcal{M}_{N+1}$ corresponding to the inelastic scattering $n\rightarrow m+1$ involving an additional outgoing graviton with momentum $k=\omega(1,\hat{n})$. The sub-subleading soft graviton theorem  relates $\mathcal{M}_{N+1}$ to $\mathcal{M}_{N}$ in the limit where the graviton is soft, \textit{i.e.} when its energy $\omega$ is small and it reads \cite{Weinberg:1965nx,Cachazo:2014fwa}
\begin{align}
\label{F1}
\mathcal{M}^{\mu\nu}_{N+1}(p_i,p_f,k)=\kappa\bigg(\hat{\mathcal{S}}_{(-1)}^{\mu\nu}+\hat{\mathcal{S}}_{(0)}^{\mu\nu}+\hat{\mathcal{S}}_{(1)}^{\mu\nu}\bigg)\mathcal{M}_{N}(p_i,p_f)+\mathcal{O}(\omega^2),
\end{align}
where $\kappa=\sqrt{8\pi G_N}$ and
\begingroup
\addtolength\jot{8pt} 
\begin{align}
\label{F2}
&\hat{\mathcal{S}}_{(-1)}^{\mu\nu}=\sum_{j=1}^{N}\frac{p_j^{\mu}p_j^{\nu}}{p_j\cdot k},\\\label{F2.1}&\hat{\mathcal{S}}_{(0)}^{\mu\nu}=\sum_{j=1}^{N}\frac{p_j^{(\mu}k_{\rho}}{p_j\cdot k}\hat{J}_j^{\nu)\rho},\\ \label{F2.2}& \hat{\mathcal{S}}_{(1)}^{\mu\nu}=\frac{1}{2}\sum_{j=1}^{N}\frac{k_{\rho}k_{\sigma}}{p_j\cdot k}\hat{J}_j^{\mu\rho}\hat{J}_j^{\nu\sigma},
\end{align}
\endgroup
are the leading, subleading and sub-subleading soft factors and, for spinless particles, $\hat{J}_j^{\mu\nu}=2p_j^{[\mu}\frac{\partial}{\partial p_{j\nu}]}$\footnote{Here we define symmetrization as $A^{(\mu}B^{\nu)}=\frac{1}{2}(A^{\mu}B^{\nu}+A^{\nu}B^{\mu})$ and antisymmetrization as $A^{[\mu}B^{\nu]}=\frac{1}{2}(A^{\mu}B^{\nu}-A^{\nu}B^{\mu})$.} is the mechanical angular momentum operator of the $j$-th particle. Equation \eqref{F1} is an analytic series expansion of $\mathcal{M}^{\mu\nu}_{N+1}$ for small $\omega$. A similar result holds for scattering of particles mediated by the electromagnetic field. 

The leading and subleading soft factors capture the full $\mathcal{O}(\omega^{-1})$ and $\mathcal{O}(\omega^{0})$  behaviours of $\mathcal{M}^{\mu\nu}_{N+1}$ and they are universal, \textit{i.e.} they do not depend on the particular theory considered. They are a direct consequence of gauge invariance of the amplitude \cite{Bern:2014vva,Broedel:2014fsa} and are related to the conservation of total momentum and angular momentum during the scattering process \cite{Weinberg:1964ew}. Further, it has been proven that they are connected to the invariance of the gravitational S-matrix under the action of asymptotic symmetries in four spacetime dimensions and to the gravitational memory effect (see, \textit{e.g.} \cite{Strominger:2017zoo}). On the other hand, the $\mathcal{O}(\omega)$ behaviour is not universal and the sub-subleading soft factor in \eqref{F2} captures only its universal part \cite{Laddha:2017ygw}.

Already at tree-level, when going to position space in $D=4$, there is a breakdown of the power series expansion in $\omega$ at subleading order and the corresponding subleading soft factor contains terms that are non-analytic in $\omega$ and, in particular, they behave as $\mathcal{O}(\log\omega)$ \cite{Laddha:2018myi,Laddha:2018vbn,Sahoo:2018lxl}.  A simple derivation of this consists in explicitly computing the trajectories of particles subjected to the gravitational interaction and observing that at early and late times, as the proper time $\abs{\tau}\rightarrow\infty$, they are logarithmically divergent. As a consequence, the mechanical angular momentum $\hat{J}_i^{\mu\nu}$ of each particle diverges as $\log\abs{\tau}\simeq	\log\abs{\omega}^{-1}$ and therefore also the subleading soft factor in \eqref{F1} does so. Ultimately, this is a consequence of the long-range nature of the gravitational (and electromagnetic) interaction in $D=4$. 
Equation \eqref{F1} is still a tree-level statement. It is of interest in our analysis to consider how it gets modified when one takes into account one-loop corrections to $\mathcal{M}_{N+1}$ and $\mathcal{M}_{N}$, together with their infrared divergences.
As discussed in the seminal paper \cite{Weinberg:1965nx}, such divergences exponentiate for (any) gravity amplitude $\mathcal{M}$
\begingroup
\addtolength\jot{8pt}
\begin{align}
\label{W1}
\mathcal{M}=e^{-\mathcal{W}}\mathcal{M}^{\mathrm{I.F.}},
\end{align}
\endgroup
where I.F. stays for ``infrared finite" and where, for a general $(N+1)$-point amplitude involving massive particles, the function $\mathcal{W}$ is explicitly given by
\begin{align}
\label{W1}
\mathcal{W}=-i\frac{G_N}{2 \epsilon}\sum_{j\neq k=1}^{N+1}w_{jk},\qquad w_{jk}\simeq m_jm_k\frac{\sigma^2_{jk}-\frac{1}{2}}{\sqrt{\sigma^2_{jk}-1}}\delta_{\eta_{j}\eta_{k},1},
\end{align}
in dimensional regularization with $\epsilon=\frac{4-D}{2}$. In equation \eqref{W1} we used ``$\simeq$" to denote equality after having taken the classical limit, in which we are eventually interested. A complete formula for $\mathcal{W}$, including its quantum contributions, can be found in appendix \ref{C}.  If we take particle $N+1$ to be outgoing and massless with momentum $k$ we get \cite{Georgoudis:2023ozp}
\begin{align}
\label{W2}
\mathcal{W}\simeq-i\frac{G_N}{2 \epsilon}\sum_{j\neq k=1}^Nw_{jk}+i\frac{G_N}{\epsilon}\sum_{j=1}^N
(p_j\cdot k)\delta_{\eta_{j},\eta_5}\equiv \mathcal{W}_{\mathrm{reg}}+\mathcal{W}_{\mathrm{phase}}.
\end{align}
As discussed later in section \ref{resum} in \eqref{conv},
$\epsilon^{-1}$ terms correspond to logarithmic terms \footnote{See also the discussion in \cite{Agrawal:2023zea}.}. In particular \cite{Krishna:2023fxg}
\begingroup
\addtolength\jot{8pt} %
\begin{align}
\label{W3}
&\mathcal{W}_{\mathrm{reg}}\simeq- iG_Nm_jm_k\sum_{j\neq k=1}^N\frac{\sigma_{jk}^2-\frac{1}{2}}{\sqrt{\sigma^2_{jk}-1}}\delta_{\eta_{j}\eta_{k},1}\log\{(\omega+i\eta_j\epsilon )L\},\\
\label{F4.a}
&\mathcal{W}_{\mathrm{phase}}\simeq 2iG_N\log\{(\omega+i\epsilon) R\}\sum_{j=n+1}^m p_j\cdot k,
\end{align}
\endgroup
where the $\epsilon$ appearing in the last two equations \textit{is not} the dimensional regularisation parameter we introduced earlier, but comes from performing loop integrals and by keeping track of the $i\epsilon$ prescription, see \textit{e.g.} appendix A of \cite{Saha:2019tub}. While in many references the $i\epsilon$ prescription inside the argument of the logarithm is not taken into account, we find it a necessary ingredient to obtain part of the result, as we explain later in section \ref{eiksoft}. $\mathcal{W}_{\mathrm{reg}}$ comes from loop diagrams where the internal graviton line connects two massive lines and $\L^{-1}=\Lambda$ is a UV scale that regularises such loop integrals. Hence, $\mathcal{W}_{\mathrm{reg}}$ factors out IR divergences contained both in $\mathcal{M}_{N+1}$ and $\mathcal{M}_{N}$ and for the latter we can write a relation
\begingroup
\addtolength\jot{10pt} 
\begin{align}
\label{F4.1}
&\mathcal{M}_N(p_i,p_f)=e^{-\mathcal{W}_{\mathrm{reg}}}\mathcal{M}^{\mathrm{I.F.}}_N(p_i,p_f),
\end{align}
\endgroup
On the other hand, $\mathcal{W}_{\mathrm{phase}}$ comes from loop diagrams involving an internal three-graviton vertex and $R$ is an arbitrary IR scale related to the energy resolution of the detector. Therefore we write
\begingroup
\addtolength\jot{8pt} %
\begin{align}
 \mathcal{M}^{\mu\nu}_{N+1}(p_i,p_f,k)=e^{-\mathcal{W}_{\mathrm{reg}}-\mathcal{W}_{\mathrm{phase}}}\mathcal{M}^{\mu\nu\hspace{0.1cm}\mathrm{I.F.}}_{N+1}(p_i,p_f,k).
\end{align}
\endgroup
Using the previous relations together with soft theorems gives
\begingroup
\addtolength\jot{9pt} %
\begin{align}
\label{F4.2}
\nonumber\mathcal{M}^{\mu\nu}_{N+1}(p_i,p_f,k)&=e^{-\mathcal{W}_{\mathrm{reg}}-\mathcal{W}_{\mathrm{phase}}}\mathcal{M}^{\mu\nu\hspace{0.1cm}\mathrm{I.F.}}_{N+1}(p_i,p_f,k)=e^{-\mathcal{W}_{\mathrm{phase}}-\mathcal{W}_{\mathrm{reg}}}\hat{\mathcal{S}}^{\mu\nu}\mathcal{M}^{\mathrm{I.F.}}_N(p_i,p_f)\\&=e^{-\mathcal{W}_{\mathrm{phase}}}e^{-\mathcal{W}_{\mathrm{reg}}}\hat{\mathcal{S}}^{\mu\nu}e^{\mathcal{W}_{\mathrm{reg}}}\mathcal{M}_N(p_i,p_f),
\end{align}
\endgroup
where
\begingroup
\addtolength\jot{4pt} %
\begin{align}
\label{F4.1}
\hat{\mathcal{S}}^{\mu\nu}=\kappa\bigg(\hat{\mathcal{S}}^{\mu\nu}_{(-1)}+\hat{\mathcal{S}}^{\mu\nu}_{(0)}+\hat{\mathcal{S}}^{\mu\nu}_{(1)}\bigg)+\mathcal{O}(\omega^2).
\end{align} 
\endgroup
is the soft operator. We get a concise formulation of the one-loop soft graviton theorem
\begingroup
\addtolength\jot{4pt} %
\begin{align}
\label{F3}
\nonumber\mathcal{M}^{\mu\nu}_{N+1}(p_i,p_f,k)=&\kappa e^{-\mathcal{W}_{\mathrm{phase}}}e^{-\mathcal{W}_{\mathrm{reg}}}\bigg(\hat{\mathcal{S}}^{\mu\nu}_{(-1)}+\hat{\mathcal{S}}^{\mu\nu}_{(0)}+\hat{\mathcal{S}}^{\mu\nu}_{(1)}\bigg)e^{\mathcal{W}_{\mathrm{reg}}}\mathcal{M}_{N}(p_i,p_f)	\\&+\mathcal{O}(\omega^2,\omega^2\log\omega).
\end{align}
\endgroup Equation \eqref{F3} can be expanded as
\begingroup
\addtolength\jot{9pt} %
\begin{align}
\label{F5}
\nonumber\mathcal{M}^{\mu\nu}_{N+1}(p_i,p_f,k)=&\kappa e^{-\mathcal{W}_{\mathrm{phase}}}\bigg\{\hat{\mathcal{S}}^{\mu\nu}_{(-1)}+\hat{\mathcal{S}}^{\mu\nu}_{(0)}+\hat{\mathcal{S}}^{\mu\nu}_{(1)}+(\hat{\mathcal{S}}_{(0)}^{\mu\nu}\mathcal{W}_{\mathrm{reg}})\\\nonumber&+\sum_{j=1}^N\frac{k_{\rho}k_{\sigma}}{p_j\cdot k}(\hat{J}_j^{\mu\rho}\mathcal{W}_{\mathrm{reg}})\hat{J}_j^{\nu\sigma}+\frac{1}{2}\sum_{j=1}^N\frac{k_{\rho}k_{\sigma}}{p_j\cdot k}(\hat{J}_j^{\mu\rho}\mathcal{W}_{\mathrm{reg}})(\hat{J}_j^{\nu\sigma}\mathcal{W}_{\mathrm{reg}})\\&+(\hat{\mathcal{S}}_{(1)}^{\mu\nu}\mathcal{W}_{\mathrm{reg}})\bigg\}\mathcal{M}_{N}(p_i,p_f)+\mathcal{O}(\omega^2,\omega^2\log\omega),
\end{align}
\endgroup
and, at one-loop one has to expand the overall exponential up to quadratic order and keep only terms of $\mathcal{O}(G_N^n)$ with $n< 2$. This procedure reproduces the results in \cite{Bern:2014oka} and in \cite{Krishna:2023fxg}. We notice that in the latter reference, the last term is absent, while it is displayed in the former. As we discuss later, we find that this term is subleading in $\hbar$ and therefore it does not contribute to the classical limit.

\section{Eikonal exponentiation revisited}
\label{eikonal}
Here we specify to the case $n=m=2$ and we consider an elastic $2\rightarrow 2$ scattering. We can go to impact parameter space by means of a Fourier transform
\begingroup
\addtolength\jot{8pt} %
\begin{align}
\label{F7}
\nonumber\tilde{\mathcal{M}}_4(s,b)&=\int \frac{d^{4-2\epsilon}q}{(2\pi)^{4-2\epsilon}}e^{ib\cdot q}\hat{\delta}(2p_1\cdot q)\hat{\delta}(2p_2\cdot q)\mathcal{M}_4(s,q^2)\\&=\frac{1}{4m_1m_2\sqrt{\sigma^2-1}}\int \frac{d^{2-2\epsilon}q_{\perp}}{(2\pi)^{2-2\epsilon}}e^{i b\cdot q_{\perp}}\mathcal{M}_4(s,q_{\perp}^2),
\end{align}
\endgroup
where we introduced an infrared regulator $\epsilon=(4-D)/2$, $\hat{\delta}(x)=2\pi\delta(x)$ and $q_{\perp}$ is the projection of $q$ onto the two-dimensional plane perpendicular to $p_1$ and $p_2$
\begingroup
\addtolength\jot{8pt} %
\begin{align}
\label{F7.1}
q^{\mu}_{\perp}=P_{\perp}^{\mu}{}_{\nu}q^{\nu},\quad P_{\perp}^{\mu\nu}=\eta^{\mu\nu}-P_{\parallel}^{\mu\nu},\quad P_{\parallel}^{\mu\nu}=\frac{1}{\sigma^2-1}\bigg(\frac{p_1^{\mu}p_1^{\nu}}{m_1^2}+\frac{p_2^{\mu}p_2^{\nu}}{m_2^2}-2\sigma\frac{p_1^{(\mu}p_2^{\nu)}}{m_1m_2}\bigg).
\end{align}
\endgroup
In \eqref{F7}  $s=-(p_1+p_2)^2=m_1^2+m_2^2+2m_1m_2\sigma$, where $\sigma\equiv\sigma_{12}=\sigma_{34}=(1-v^2)^{-\frac{1}{2}}$, $v$ being the relative velocity between the two particles and $t=-(p_1+p_4)^2=-q^2$ are the usual Mandelstam variables. In the PM expansion we assume that the Schwarzschild radius $r_j=2G_Nm_j$ of the two particles is much smaller than the impact parameter $b$ of the scattering process. This allows us to expand perturbatively $\mathcal{M}_4$ as
\begin{align}
\label{F7.2}
\mathcal{M}_4(s,q^2)=\sum_{n\geq 0}\mathcal{M}^{(n)}_4(s,q^2),
\end{align}
where $\mathcal{M}^{(n)}_4\sim G^{n+1}_N$ is the $n$-loop ($n+1$ PM) contribution to the amplitude arising from the exchange of $n+1$ virtual gravitons. As discussed \textit{e.g.} in \cite{DiVecchia:2023frv} in the classical limit the amplitude in impact parameter space resums into a compact exponential structure \footnote{At 3PM the eikonal develops an imaginary part due to the emission of gravitational radiation. In this case the eikonal should be more properly treated as an operator, see Ref. \cite{DiVecchia:2022nna}.}
\begin{align}
\label{F7.3}
1+i\tilde{\mathcal{M}}_4(s,b)\simeq e^{2i\delta(\sigma,b)},\qquad \delta(\sigma,b)=\sum_{n\geq 0}\delta^{(n)}(\sigma,b),
\end{align}
where $\delta$ is the eikonal phase and $\delta^{(n)}$ its $n$-loop contribution that scales as
\begin{align}
\label{F7.4}
\delta^{(n)}(\sigma,b)\sim (G_N s)\bigg(\frac{G_N \sqrt{s}}{b}\bigg)^n,
\end{align}
For instance, the tree-level (1PM) amplitude corresponding to one-graviton exchange is
\begin{align}
\label{F6}
i\mathcal{M}^{(0)}_4(s,q)=\vcenter{\hbox{\includegraphics[width=4.5cm,height=4.4cm]{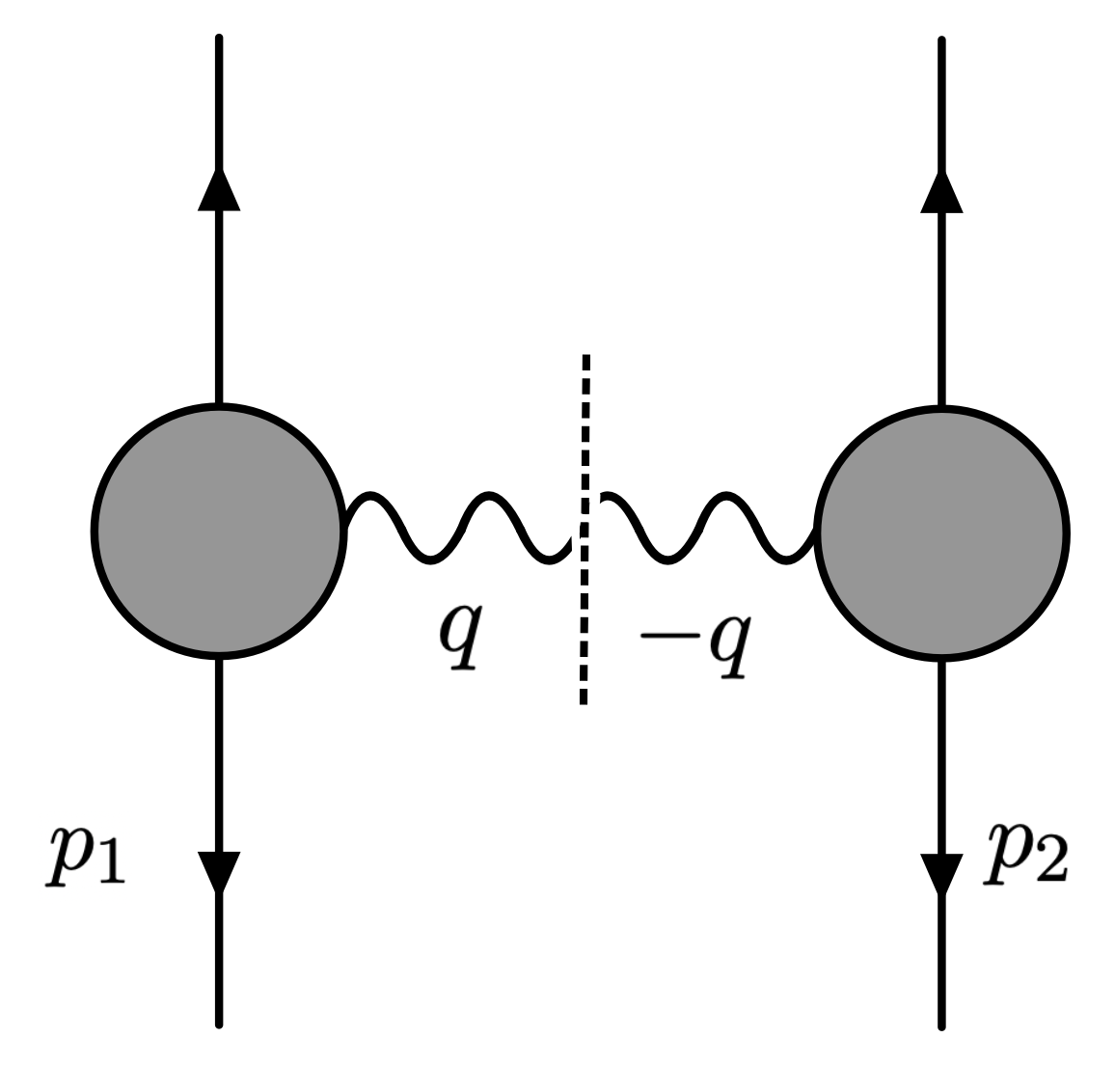}}}=\frac{4i\kappa^2}{q^2}m_1^2m_2^2\left(\sigma^2-\frac{1}{D-2}\right)+\mathcal{O}((q^2)^0).
\end{align}
In this case the integral in \eqref{F7} is infrared divergent and it gives \cite{KoemansCollado:2019ggb,Cristofoli:2020uzm}
\begin{align}
\label{F8}
2i\delta^{(0)}(\sigma,b)=i\tilde{\mathcal{M}}^{(0)}_4(s,b)=iG_Nm_1m_2\frac{2\sigma^2-1}{\sqrt{\sigma^2-1}}\frac{\Gamma(-\epsilon)}{(\pi b^2)^{-\epsilon}}.
\end{align}
Such infrared divergence can be traced back to the usual Coulombic potential in three space dimensions. In this case, introducing an inverse length scale $\mu$, the leading eikonal becomes
\begin{align}
\label{F9}
2\delta^{(0)}(\sigma,b)=-2G_Nm_1m_2\frac{2\sigma^2-1}{\sqrt{\sigma^2-1}}\log(\mu b).
\end{align}
The one-loop eikonal is not infrared divergent and it is given by
\begin{align}
\label{F10}
2\delta^{(1)}(\sigma,b)= \frac{3\pi G^2_Nm_1m_2(m_1+m_2)}{4b}\frac{5\sigma^2-1}{\sqrt{\sigma^2-1}}.
\end{align}
After the classical exponentiation in \eqref{F7.3}, one can perform an inverse Fourier transform to get the classical amplitude in momentum space
\begin{align}
\label{F11}
\mathcal{M}_4(s,Q)=4im_1m_2\sqrt{\sigma^2-1}\int d^2b\hspace{0.1cm}
e^{-iQ\cdot b}\bigg(1-e^{2i\delta(\sigma,b)}\bigg).
\end{align}
Here $Q$ is the classical transfer momentum, that is different from $q$, which scales with $\hbar$. By the usual argument \cite{DiVecchia:2023frv}, the integral is dominated by the stationary phase and therefore we identify the total classical momentum transfer as
\begin{align}
\label{F12}
\nonumber Q^{\mu}&=\frac{\partial\hspace{0.05cm}}{\partial b_{\mu}} \mathrm{Re}\hspace{0.05cm}2\delta(\sigma,b)=Q^{\mu}_{(0)}+Q^{\mu}_{(1)}+\mathcal{O}(G^3_N)\\&=-2G_Nm_1m_2\frac{2\sigma^2-1}{\sqrt{\sigma^2-1}}\frac{b^{\mu}}{b^2}+\frac{3\pi G^2_Nm_1m_2(m_1+m_2)}{4}\frac{5\sigma^2-1}{\sqrt{\sigma^2-1}}\frac{b^{\mu}}{b^3}+\mathcal{O}(G_N^3).
\end{align}

\section{Soft expansion in $b$ space}
In this section we discuss the soft expansion of the classical amplitude in impact parameter space. We define a Fourier transform for the 5-point amplitude, expand the integrand for small frequencies, and then compute the integral collecting terms  that scale with the same power of $\omega$.
\label{resum}
\subsection*{Setup}
Let us consider the Fourier transform of the classical 5-point amplitude, which we denote by $\tilde{\mathcal{M}}^{\mu\nu}_{5}$. Introducing the transferred momenta  $q_1=p_1+p_4$ and $q_2=p_2+p_3$, it can be written as
\begin{align}
\label{F157}
\nonumber
\tilde{\mathcal{M}}^{\mu\nu}_{5}(p_i,b_i,k)=&\int\frac{d^D q_1}{(2\pi)^D}\int\frac{d^D q_2}{(2\pi)^D}e^{i(b_1\cdot q_1+b_2\cdot q_2)}\mathcal{M}^{\mu\nu}_{5}(p_i,q_i,k)\hat{\delta}(2p_1\cdot q_1)\hat{\delta}(2p_2\cdot q_2)\\&\times\hat{\delta}^{(D)}(q_1+q_2+k).
\end{align}
Solving the $D$-dimensional $\hat{\delta}$ by writing $q_1= q- \frac{k}{2}$ and $q_2 = - \frac{k}{2} -q$, gives 
\begin{align}
\label{F160}
\nonumber\tilde{\mathcal{M}}^{\mu\nu}_{5}(p_i,b_i,k)=&e^{-i\frac{(b_1+b_2)\cdot k}{2}}\int\frac{d^D q}{(2\pi)^D}e^{i b\cdot q}\mathcal{M}^{\mu\nu}_{5}\bigg(p_i,,q-\frac{k}{2},-q-\frac{k}{2},k\bigg)\\&\times\hat{\delta}(2p_1\cdot q-p_1\cdot k)\hat{\delta}(2p_2\cdot q+p_2\cdot k),
\end{align}
where we denoted $b\equiv b_1-b_2$.
Now we proceed to solve the remaining two $\hat{\delta}$ functions in a covariant way. We start by noticing that any four-vector $v$ can be decomposed into a component that lies in the plane spanned by $p_1$ and $p_2$ and into a component perpendicular to it
\begin{align}
\label{F165}
v^{\mu}=v^{\mu}_{\parallel}+v^{\mu}_{\perp},\qquad v^{\mu}_{\parallel}=P_{\parallel}^{\mu\nu}v_{\nu},\qquad v^{\mu}_{\perp}=P_{\perp}^{\mu\nu}v_{\nu},
\end{align}
where the parallel and perpendicular projectors have been defined in \eqref{F7.1}. The arguments of the two delta functions appearing in the Fourier transform \eqref{F160} can be written as
\begin{align}
\label{F166}
2p_1\cdot q-p_1\cdot k=2p_1\cdot q_{\parallel}-p_1\cdot k_{\parallel},\qquad2p_2\cdot q+p_2\cdot k=2p_2\cdot q_{\parallel}+p_2\cdot k_{\parallel}.
\end{align}
Furthermore we can decompose $v_{\parallel}$ as
\begin{align}
\label{F167}
v^{\mu}_{\parallel}=-v_{1\parallel}\frac{p_1^{\mu}}{m_1}-v_{2\parallel}\frac{p_2^{\mu}}{m_2},
\end{align}
with
\begin{align}
v_{1\parallel}=\frac{\sigma(v\cdot p_2) m_1-(v\cdot p_1)m_2}{m_1m_2(\sigma^2-1)},\qquad v_{2\parallel}=\frac{\sigma(v\cdot p_1) m_2-(v\cdot p_2)m_1}{m_1m_2(\sigma^2-1)}.
\end{align}
Solving the first of \eqref{F166} gives
\begin{align}
\label{F168}
q_{1\parallel}^*=\frac{k_{1\parallel}+\sigma k_{2\parallel}-2\sigma q_{2\parallel}}{2},
\end{align}
and the $\hat{\delta}$ is
\begin{align}
\label{F168.a}
\hat{\delta}(2p_1\cdot q-p_1\cdot k)=\frac{1}{2m_1}\hat{\delta}(q_{1\parallel}-q_{1\parallel}^*).
\end{align}
Solving the second of \eqref{F166} gives, using \eqref{F168.a}
\begin{align}
\label{F168.b}
q_{2\parallel}^*=\frac{k_{2\parallel}(\sigma^2+1)+2k_{1\parallel}\sigma}{2(\sigma^2-1)},
\end{align}
and the $\hat{\delta}$ becomes
\begin{align}
\label{F169.c}
\hat{\delta}(2p_2\cdot q+p_2\cdot k)=\frac{1}{2m_2(\sigma^2-1)}\hat{\delta}(q_{2\parallel}-q_{2\parallel}^*).
\end{align}
Putting \eqref{F168.b} back into \eqref{F168} gives
\begin{align}
\label{F169}
q_{1\parallel}^*=\frac{k_{1\parallel}(\sigma^2+1)+2k_{2\parallel}\sigma}{2(1-\sigma^2)}.
\end{align}
Note that, for $k=0$, we get $q_{1\parallel}^*=0=q_{2\parallel}^*$, as expected, and the kinematics of the 5-point amplitude collapses into the one of the elastic $2\rightarrow 2$ amplitude. Let us now compute the transformation of the measure $d^{D}q$ appearing in \eqref{F160}. We are splitting the integration variables $q$ as $q^{\mu}=\{q_{\parallel}^{\alpha},q_{\perp}^{i}\}=\{q_{1\parallel},q_{2\parallel},q^i_{\perp}\}$ with $\alpha=1,2$ and $i=1,...,D-2$. We have $d^Dq=d^{D-2}q_{\perp}d^2q_{\parallel}$ and, using the standard center of mass parametrization
\begin{align}
\label{F15.1}
p_1^{\mu}=-(E_1,0,p,0),\qquad p_2^{\mu}=-(E_2,0,-p,0),
\end{align}
with
\begin{align}
\label{F15.2}
E_1=\frac{m_1(m_1+m_2\sigma)}{\sqrt{m_1^2+m_2^2+2\sigma m_1m_2}},\qquad E_2=\frac{m_2(m_2+m_1\sigma)}{\sqrt{m_1^2+m_2^2+2\sigma m_1m_2}},
\end{align}
\begin{align}
\label{F15.3}
p=\frac{m_1m_2\sqrt{\sigma^2-1}}{\sqrt{m_1^2+m_2^2+2m_1m_2\sigma}},
\end{align}
we have $q_{\parallel}^{\alpha}=(q^0,q^3)$ and hence the coefficients $q_{1\parallel}$ and $q_{2\parallel}$ of the decomposition \eqref{F167} are explicitly given by
\begin{align}
\label{F170}
q^0=q_{1\parallel}\frac{E_1}{m_1}+q_{2\parallel}\frac{E_2}{m_2},\qquad q^3=q_{1\parallel}\frac{p}{m_1}-q_{2\parallel}\frac{p}{m_2},
\end{align}
and hence $
dq^0dq^3=\sqrt{\sigma^2-1}dq_{\parallel1}dq_{\parallel2}$ so that
\begingroup
\addtolength\jot{9pt}
\begin{align}
\label{F172}
\nonumber \tilde{\mathcal{M}}^{\mu\nu}_{5}(p_i,b_i,k)=&\frac{e^{-i\frac{(b_1+b_2)\cdot k}{2}}}{4m_1m_2\sqrt{\sigma^2-1}}\int \frac{d^{D-2}q_{\perp}
}{(2\pi)^{D-2}}\frac{dq_{1\parallel }dq_{2\parallel }}{(2\pi)^2}e^{ib\cdot q}\hat{\delta}(q_{1\parallel}-q_{1\parallel}^*)\hat{\delta}(q_{2\parallel}-q_{2\parallel}^*)\\&\nonumber\times\mathcal{M}^{\mu\nu}_{5}\bigg(p_i,q-\frac{k}{2},-q-\frac{k}{2},k\bigg)\\=&\frac{e^{-i\frac{(b_1+b_2)\cdot k}{2}}}{4m_1m_2\sqrt{\sigma^2-1}}\int\frac{d^{2-2\epsilon}q_{\perp}}{(2\pi)^{2-2\epsilon}}e^{ib\cdot q_{\perp}}\mathcal{M}^{\mu\nu}_{5}\bigg(p_i,q-\frac{k}{2},-q-\frac{k}{2},k\bigg)\bigg|_{q_{1\parallel }^*,q_{2\parallel }^*}.
\end{align}
\endgroup
We notice that the Fourier transform of the classical 5-point amplitude is, apart from the overall exponential (which can be however set to zero by choosing a frame where $b_1=-b_2$), is the same as the one involving the elastic amplitude in \eqref{F7}, but the integrand does not have to be evaluated in $q_{\perp}$ but in $(q_{i\parallel }^*,q_{\perp})$, where $q_{i \parallel}^*$ depend linearly on $k\sim \omega$.

In the following we use equation \eqref{F172} to derive the soft expansion of the tree-level 5-point amplitude. 
\subsection*{Classical 5-point amplitude and its soft expansion}
An explicit expression of the classical tree-level 5-point amplitude is \cite{Luna:2017dtq,Mougiakakos:2021ckm,Manohar:2022dea,Goldberger:2016iau,Jakobsen:2021lvp}
\begingroup
\addtolength\jot{9pt}
\begin{align}
\label{clam}
\nonumber &\mathcal{M}^{(0)\mu\nu}_{5}(p_i,q_i,k)=\kappa^{3}\Bigg\{ 8\frac{(p_1\cdot kp_2^{\mu}-p_2\cdot kp_1^{\mu})(p_1\cdot kp_2^{\nu}-p_2\cdot kp_1^{\nu})}{q_1^2q_2^2}\\\nonumber+& 8 p_1\cdot p_2\left[\frac{\frac{p_2\cdot k}{p_1\cdot k}p_1^{\mu}p_1^{\nu}-p_1^{(\mu}p_2^{\nu)}}{q_2^2}+\frac{\frac{p_1\cdot k}{p_2\cdot k}p_2^{\mu}p_2^{\nu}-p_1^{(\mu}p_2^{\nu)}}{q_1^2}-2\frac{p_1\cdot k p_2^{(\mu}q_1^{\nu)}-p_2\cdot k p_1^{(\mu}q_1^{\nu)}}{q_1^2q_2^2}\right]\\+&\beta\left[-\frac{(q_1\cdot k)}{(p_1\cdot k)^2q_2^2}p_1^{\mu}p_1^{\nu}-\frac{(q_2\cdot k)}{(p_2\cdot k)^2q_1^2}p_2^{\mu}p_2^{\nu}+2\left(\frac{p_1^{(\mu}q_1^{\nu)}}{(p_1\cdot k)q_2^2}-\frac{p_2^{(\mu}q_1^{\nu)}}{(p_2\cdot k)q_1^2}+\frac{q_1^{\mu}q_1^{\nu}}{q_1^2q_2^2}\right)\right]\Bigg\},
\end{align}
\endgroup
with $\beta=4m_1^2m_2^2(\sigma^2-\frac{1}{D-2})$ and $q_1+q_2+k=0$. Notice that the amplitude in \eqref{clam} is gauge invariant, $k_{\mu}\mathcal{M}_{5}^{(0)\mu\nu}=k_{\nu}\mathcal{M}_{5}^{(0)\mu\nu}=0$. Using $q_1=q-\frac{k}{2}$ and $q_2=-q-\frac{k}{2}$, we can expand the previous amplitude around $k=0$ up to the linear order in $\omega$. We get
\begingroup
\addtolength\jot{9pt}
\begin{align}
\label{F103.a}
\nonumber\mathcal{M}^{(0)\mu\nu}_{5}\bigg(p_i,q-\frac{k}{2},-q-\frac{k}{2},k\bigg)=&\left.\mathcal{M}_{5}^{(0)\mu\nu}\right|_{\mathcal{O}(\omega^{-1})}(p_i,q,k)+\left.\mathcal{M}_{5}^{(0)\mu\nu}\right|_{\mathcal{O}(\omega^{0})}(p_i,q,k)\\&+\left.\mathcal{M}_{5}^{(0)\mu\nu}\right|_{\mathcal{O}(\omega)}(p_i,q,k)+\mathcal{O}(\omega^2),
\end{align}
\endgroup
with
\begingroup
\addtolength\jot{9pt}
\begin{align}
\label{F103}
    \nonumber&\left.\mathcal{M}_{5}^{(0)\mu\nu}\right|_{\mathcal{O}(\omega^{-1})}(p_i,q,k)=-\frac{\kappa^{3}\beta}{q^2}\bigg[\bigg(\frac{p_1^{\mu}p_1^{\nu}}{(p_1\cdot k)^2}-\frac{p_2^{\mu}p_2^{\nu}}{(p_2\cdot k)^2}\bigg)(q\cdot k)-2\bigg(\frac{p_1^{(\mu}q^{\nu)}}{p_1\cdot k}\\&-\frac{p_2^{(\mu}q^{\nu)}}{p_2\cdot k}\bigg)\bigg],\\
\label{subl}\nonumber&\left.\mathcal{M}_{5}^{(0)\mu\nu}\right|_{\mathcal{O}(\omega^{0})}(p_i,q,k)=\kappa^{3}\Bigg\{8p_1\cdot p_2\frac{\frac{p_2\cdot k}{p_1\cdot k}p_1^{\mu}p_1^{\nu}+\frac{p_1\cdot k}{p_2\cdot k}p_2^{\mu}p_2^{\nu}-2p_1^{(\mu}p_2^{\nu)}}{q^2}+\\&+\beta\bigg[\frac{(q\cdot k)^2}{q^4}\bigg(\frac{p_1^{\mu}p_1^{\nu}}{(p_1\cdot k)^2}+\frac{p_2^{\mu}p_2^{\nu}}{(p_2\cdot k)^2}\bigg)-2\frac{(q\cdot k)}{q^4}\bigg(\frac{p_1^{(\mu}q^{\nu)}}{p_1\cdot k}+\frac{p_2^{(\mu}q^{\nu)}}{p_2\cdot k}\bigg)+2\frac{q^{\mu}q^{\nu}}{q^4}\bigg]\Bigg\}, 
\end{align}
\endgroup
\begingroup
\addtolength\jot{9pt}
\begin{align}
\nonumber &\left.\mathcal{M}_{5}^{(0)\mu\nu}\right|_{\mathcal{O}(\omega)}(p_i,q,k)=\kappa^{3}\left\{8\frac{[(p_1\cdot k)p_2^{\nu}-(p_2\cdot k)p_1^{\nu}][p_1\cdot k)p_2^{\mu}-(p_2\cdot k)p_1^{\mu}]}{q^4}\right.\\\nonumber &+8p_1\cdot p_2\bigg[\frac{\frac{p_1\cdot k}{p_2\cdot k}p_2^{\mu}p_2^{\nu}-\frac{p_2\cdot k}{p_1\cdot k}p_1^{\mu}p_1^{\nu}}{q^4}(q\cdot k)-2\frac{(p_1\cdot k) p_2^{(\mu}q^{\nu)}-(p_2\cdot k) p_1^{(\mu}q^{\nu)}}{q^4}\bigg]\\&-\beta\bigg[\bigg(\frac{p_1^{\mu}p_1^{\nu}}{(p_1\cdot k)^2}-\frac{p_2^{\mu}p_2^{\nu}}{(p_2\cdot k)^2}\bigg)\frac{(q\cdot k)^3}{q^6}-2\bigg(\frac{p_1^{(\mu}q^{\nu)}}{p_1\cdot k}-\frac{p_2^{(\mu}q^{\nu)}}{p_2\cdot k}\bigg)\frac{(q\cdot k )^2}{q^6}\bigg]\Bigg\}.
\end{align}
\endgroup
As can be easily shown, these quantities can be obtained by acting with the soft operators on the classical  tree-level, elastic amplitude displayed in \eqref{F6}, and then taking the classical limit
\begingroup
\addtolength\jot{9pt}
\begin{align}
&\mathcal{M}^{(0)\mu\nu}_{5}|_{\mathcal{O}(\omega^{-1})}(p_i,q,k)=\kappa\hat{\mathcal{S}}_{(-1)}^{\mu\nu}\mathcal{M}^{(0)}_4(p_i,q),\\&\mathcal{M}^{(0)\mu\nu}_{5}|_{\mathcal{O}(\omega^0)}(p_i,q,k)=\kappa\hat{\mathcal{S}}_{(0)}^{\mu\nu}\mathcal{M}^{(0)}_4(p_i,q),\\&\mathcal{M}^{(0)\mu\nu}_{5}|_{\mathcal{O}(\omega)}(p_i,q,k)=\kappa\hat{\mathcal{S}}_{(1)}^{\mu\nu}\mathcal{M}^{(0)}_4(p_i,q).\end{align}
\endgroup
\subsection*{Leading soft}
Let us start by considering the $\mathcal{O}(\omega^{-1})$ contribution to the tree-level waveform in \eqref{F172}. It is clear that the only contribution  is given by
\begin{align}
\tilde{\mathcal{M}}^{(0)\mu\nu}_{5}|_{\mathcal{O}(\omega^{-1})}(p_i,b_i,k)=\frac{1}{4m_1m_2\sqrt{\sigma^2-1}}\int \frac{d^{2-2\epsilon}q_{\perp}}{(2\pi)^{2-2\epsilon}}e^{ib\cdot q_{\perp}}\mathcal{M}^{(0)\mu\nu}_{5}|_{\mathcal{O}(\omega^{-1})}(p_i,q_{\perp},k).
\end{align}
In the following we will use \cite{KoemansCollado:2019ggb}
\begin{align}
\nonumber\mathcal{I}^{\mu_1\dots \mu_n}_{\epsilon}(b,\nu)&=\int\frac{d^{2-2\epsilon} q_{\perp}}{(2\pi)^{2-2\epsilon}}e^{iq_{\perp}\cdot b}\frac{q_{\perp}^{\mu_1}\dots q_{\perp}^{\mu_n}}{(q_{\perp}^2)^{\nu}}\\&=(-i)^n\frac{\partial^n}{\partial b_{\mu_1}\dots\partial b_{\mu_n}}\frac{\Gamma(1-\epsilon-\nu)}{4^{\nu}\pi^{1-\epsilon}\Gamma(\nu)}\frac{1}{(b^2)^{1-\epsilon-\nu}},
\end{align}
with \cite{Alessio:2023kgf}
\begin{align}
\label{pr3}
\frac{\partial^n}{\partial b_{\mu_1}\dots\partial b_{\mu_n}}\frac{1}{(b^2)^{1-\epsilon-\nu}}=\sum_{m=0}^{[\frac{n}{2}]}\frac{2^{n-m}\Gamma(\nu+\epsilon)}{\Gamma(1+\epsilon-n+m)}\frac{1}{(b^2)^{1-\epsilon-\nu-n+m}}\{[P_{\perp}]^m[b]^{n-2m}\}^{\mu_1\dots\mu_n},
\end{align}
where the symbol $\{[P_{\perp}]^m[b]^{n-2m}\}^{\mu_1\dots\mu_n}$ comprises all the symmetric structures one can construct with $P_{\perp}$ and $b$ having $n$ covariant indices, \textit{e.g.}
\begingroup
\addtolength\jot{9pt}
\begin{align}
\label{pr4}
&\{[P_{\perp}]^0[b]^1\}^{\mu_1}=b^{\mu_1},\\
&\{[P_{\perp}]^1[b]^0\}^{\mu_1\mu_2}=P_{\perp}^{\mu_1\mu_2},\\
&\{[P_{\perp}]^1[b]^1\}^{\mu_1\mu_2\mu_3}=P_{\perp}^{\mu_1\mu_2}b^{\mu_3}+P_{\perp}^{\mu_2\mu_3}b^{\mu_1}+P_{\perp}^{\mu_3\mu_1}b^{\mu_2}.
\end{align}
\endgroup
We get
\begin{align}
    \nonumber \tilde{\mathcal{M}}^{(0)\mu\nu}_{5}|_{\mathcal{O}(\omega^{-1})}(p_i,b_i,k)=&-2i\kappa\frac{ G_N m_1m_2}{b^2}\frac{(2\sigma^2-1)}{\sqrt{\sigma^2-1}}\bigg[\bigg(\frac{p_1^{\mu}p_1^{\nu}}{(p_1\cdot k)^2}-\frac{p_2^{\mu}p_2^{\nu}}{(p_2\cdot k)^2}\bigg)(b\cdot k)\\&-2\bigg(\frac{p_1^{(\mu}b^{\nu)}}{p_1\cdot k}-\frac{p_2^{(\mu}b^{\nu)}}{p_2\cdot k}\bigg)\bigg],
\end{align}
reproducing the linear memory  term.
\subsection*{Subleading soft}
The $\mathcal{O}(\omega^0)$ contribution to the waveform is more subtle to construct. In fact, it is not only given by the Fourier transform of $\mathcal{M}^{(0)\mu\nu}_{5}|_{\mathcal{O}(\omega)}$ in equation \eqref{subl}, but there are also two additional contributions. The first comes from the $\mathcal{O}(\omega^0)$ term of the expansion of the leading soft factor computed in $q=(q_{\parallel}^*(\omega),q_{\perp})$:
\begin{align}
\nonumber&\left.\mathcal{M}_{5}^{(0)\mu\nu}\right|_{\mathcal{O}(\omega^{-1})}(p_i,q_{\parallel}^*,q_{\perp}k)=-\frac{\kappa^{3}\beta}{q^2_{\perp}+q_{\parallel}^{*2}}\bigg[\bigg(\frac{p_1^{\mu}p_1^{\nu}}{(p_1\cdot k)^2}-\frac{p_2^{\mu}p_2^{\nu}}{(p_2\cdot k)^2}\bigg)( q_{\perp}\cdot k+ q_{\parallel}^*\cdot k)\\\nonumber&-2\bigg(\frac{p_1^{(\mu}(q^{\nu)}_{\perp}+q_{\parallel}^{*\nu)})}{p_1\cdot k}-\frac{p_2^{(\mu}(q^{\nu)}_{\perp}+q_{\parallel}^{*\nu)})}{p_2\cdot k}\bigg)\bigg]\\&\stackrel{\mathcal{O}(\omega^0)}{=}-\frac{\kappa^{3}\beta}{q^2_{\perp}}\bigg[\bigg(\frac{p_1^{\mu}p_1^{\nu}}{(p_1\cdot k)^2}-\frac{p_2^{\mu}p_2^{\nu}}{(p_2\cdot k)^2}\bigg)( q_{\parallel}^*\cdot k)-2\bigg(\frac{p_1^{(\mu}q_{\parallel}^{*\nu)}}{p_1\cdot k}-\frac{p_2^{(\mu}q^{\nu)}q_{\parallel}^{*\nu)}}{p_2\cdot k}\bigg)\bigg].
\end{align}
The second, comes from the first order of expansion of the overall exponential $e^{-i\frac{(b_1+b_2)\cdot k}{2}}$ appearing in equation \eqref{F172} that multiplies the leading soft factor
\begin{align}
-i\frac{(b_1+b_2)\cdot k}{2}\mathcal{M}^{(0)\mu\nu}_{5}|_{\mathcal{O}(\omega^{-1})}(p_i,q,k),
\end{align}
where $\mathcal{M}^{(0)\mu\nu}_{5}|_{\mathcal{O}(\omega^{-1})}$ is given in equation \eqref{F103}.
Putting the above two equations together with \eqref{subl} gives a tensor
\begin{align}
\label{A1}
\nonumber&\mathcal{T}^{\mu\nu}_{(0)}(p_i,q,k)=-\frac{\kappa^3\beta}{q^2}\left[\frac{( q_{\parallel}^*\cdot k_{\parallel})}{(p_1\cdot k)^2}p_1^{\mu}p_1^{\nu}-\frac{p_2^{\mu}p_2^{\nu}(q_{\parallel}^*\cdot k_{\parallel})}{(p_2\cdot k)^2}-2\left(\frac{p_1^{(\mu}q^{*\nu)}_{\parallel}}{(p_1\cdot k)}-\frac{p_2^{(\mu}q^{*\nu)}_{\parallel}}{(p_2\cdot k)}\right)\right]\\\nonumber&+\kappa^3\Bigg\{8p_1\cdot p_2\frac{p_1^{\mu}p_1^{\nu}\frac{p_2\cdot k}{p_1\cdot k}+p_2^{\mu}p_2^{\nu}\frac{p_1\cdot k}{p_2\cdot k}-2p_1^{(\mu}p_2^{\nu)}}{q^2}+\\\nonumber&+\beta\bigg[\frac{(k\cdot q)^2}{q^4}\bigg(\frac{p_1^{\mu}p_1^{\nu}}{(p_1\cdot k)^2}+\frac{p_2^{\mu}p_2^{\nu}}{(p_2\cdot k)^2}\bigg)-2\frac{(q\cdot k)}{q^4}\bigg(\frac{p_1^{(\mu}q^{\nu)}}{(p_1\cdot k)}+\frac{p_2^{(\mu}q^{\nu)}}{(p_2\cdot k)}\bigg)+2\frac{q^{\mu}q^{\nu}}{q^4}\bigg]\Bigg\}\\&+i\frac{(b_1+b_2)\cdot k}{2}\frac{\kappa^{3}\beta}{q^2}\bigg[\bigg(\frac{p_1^{\mu}p_1^{\nu}}{(p_1\cdot k)^2}-\frac{p_2^{\mu}p_2^{\nu}}{(p_2\cdot k)^2}\bigg)(k\cdot q)-2\bigg(\frac{p_1^{(\mu}q^{\nu)}}{p_1\cdot k}-\frac{p_2^{(\mu}q^{\nu)}}{p_2\cdot k}\bigg)\bigg],
\end{align}
whose Fourier transform is the $\mathcal{O}(\omega^0)$ contributions to the waveform $\tilde{\mathcal{M}}^{(0)\mu\nu}_{5}$.  Performing such a  Fourier transform of $\mathcal{T}^{\mu\nu}_{(0)}$
gives two separate contributions. The first comes only from the last two lines and it is
\begin{align}
\label{om0}
\nonumber\tilde{\mathcal{M}}_{5}^{(0)\mu\nu}|_{\mathcal{O}(\omega^0)}(p_i,b_i,k)=&-2\kappa\frac{G_Nm_1m_2}{b^2}\frac{2\sigma^2-1}{\sqrt{\sigma^2-1}}\bigg[\bigg(\frac{b_1\cdot k}{(p_1\cdot k)^2}p_1^{\mu}p_1^{\nu}-\frac{b_2\cdot k}{(p_2\cdot k)^2}p_2^{\mu}p_2^{\nu}\bigg)(b\cdot k)\\&-2\bigg(\frac{b_1\cdot k}{p_1\cdot k}p_1^{(\mu}b^{\nu)}-\frac{b_2\cdot k}{p_2\cdot k}p_2^{(\mu}b^{\nu)}\bigg)+b^{\mu}b^{\nu}\bigg].
\end{align}
The second is infrared divergent and needs to be regularised
\begin{align}
\nonumber\tilde{\mathcal{M}}_{5}^{(0)\mu\nu}|_{\mathcal{O}(\log(\omega))}(p_i,b_i,k)=&-\kappa G_N\frac{\sigma(2\sigma^2-3)}{(\sigma^2-1)^{\frac{3}{2}}}\bigg(\frac{p_2\cdot k}{p_1\cdot k}p_1^{\mu}p_1^{\nu}+\frac{p_1\cdot k}{p_2\cdot k}p_2^{\mu}p_2^{\nu}\\&-p_1^{\mu}p_2^{\nu}-p_1^{\nu}p_2^{\mu}\bigg)\Gamma(-\epsilon)(b^2)^{\epsilon}.
\end{align}
We use the following regularisation
\begin{align}
    &\Gamma(-\epsilon)(b^2)^{\epsilon}\longrightarrow -2\log(\omega b),\end{align}
  or in other words we just replace the $\epsilon^{-1}$ pole by a logarithm as
    \begin{align}
    \label{conv}\frac{1}{2\epsilon}\longrightarrow \log(\omega b).\end{align} The appearance of $\omega $ inside the argument of the logarithm makes it dimensionless. This can be explained as follows. In fact, in the soft $k\sim \omega\rightarrow 0$ limit the divergent integrals over  $q_{\perp}$ should be more properly treated with the method of regions and sum the contributions of the integrals over the two regions defined by $\omega\ll q_{\perp}\sim b^{-1}$ and $\omega \sim q_{\perp}\ll b^{-1}$ \footnote{We thank Carlo Heissenberg for several clarifying discussions on this point.}. In the sum, the divergence cancels  and the argument of the logarithm becomes dimensionless as it should, giving eventually a term
\begin{align}
\label{sublog}
\nonumber\tilde{\mathcal{M}}_{5}^{(0)\mu\nu}|_{\mathcal{O}(\log(\omega))}(p_i,b_i,k)=&2\kappa G_N\frac{\sigma(2\sigma^2-3)}{(\sigma^2-1)^{\frac{3}{2}}}\bigg(\frac{p_2\cdot k}{p_1\cdot k}p_1^{\mu}p_1^{\nu}+\frac{p_1\cdot k}{p_2\cdot k}p_2^{\mu}p_2^{\nu}\\&-p_1^{\mu}p_2^{\nu}-p_1^{\nu}p_2^{\mu}\bigg)\log(\omega b),
\end{align}
reproducing the result in \textit{e.g.} equation (8.96) of \cite{DiVecchia:2023frv}.
However, because of the regularisation procedure, the $\mathcal{O}(\omega^0)$ term in \eqref{om0} gets modified. The full $\mathcal{O}(\omega^0)$ contribution to $\tilde{\mathcal{M}}^{(0)\mu\nu}_5$ can be found in (4.16) of \cite{Georgoudis:2023eke}.
\subsection*{Sub-subleading soft}
Similarly, we could derive the sub-subleading contribution to the waveform. We find again a term proportional to $\omega$
\begin{align}
\label{subsub1}
\nonumber&\tilde{\mathcal{M}}^{(0)\mu\nu}_5|_{\mathcal{O}(\omega)}(p_i,b_i,k)=i\kappa \frac{G_N m_1m_2}{b^2}\frac{2\sigma^2-1}{\sqrt{\sigma^2-1}}\bigg[(b\cdot k)\bigg(\frac{(b_1\cdot k)^2}{(p_1\cdot k)^2}p_1^{\mu}p_1^{\nu}-\frac{(b_2\cdot k)^2}{(p_2\cdot k)^2}p_2^{\mu}p_2^{\nu}\bigg)\\&+2\bigg((b_1\cdot k)\frac{p_1^{(\mu}b^{\nu)}}{p_1\cdot k}+(b_2\cdot k)\frac{p_2^{(\mu}b^{\nu)}}{p_2\cdot k}\bigg)+b^{\mu}b^{\nu}\bigg],
\end{align}
which gets however modified by  the regularisation procedure for the $\omega\log(\omega b)$ term, which instead is universal  and it reads
\begin{align}
\nonumber
\label{subsublog}
&\tilde{\mathcal{M}}^{(0)\mu\nu}_5|_{\mathcal{O}(\omega\log(\omega))}(p_i,b_i,k)=-2i\kappa G_N\frac{\sigma(2\sigma^2-3)}{(\sigma^2-1)^{\frac{3}{2}}}\bigg[(b_1\cdot k)\frac{p_2\cdot k}{p_1\cdot k}p_1^{\mu}p_1^{\nu}+(b_2\cdot k)\frac{p_1\cdot k}{p_2\cdot k}p_2^{\mu}p_2^{\nu}\\&-(b_1\cdot k+b_2\cdot k)p_1^{(\mu}p_2^{\nu)}-(p_2\cdot k)p_1^{(\mu}b^{\nu)}+(p_1\cdot k)p_2^{(\mu}b^{\nu)}\bigg]\log(\omega b).
\end{align}
The last term is also in agreement with \textit{e.g.} equation (4.13) of \cite{Georgoudis:2023eke}. It also agrees with \cite{Ghosh:2021bam}. 

We notice that equations \eqref{sublog} and \eqref{subsublog} have a factor $\frac{\sigma(2\sigma^2-3)}{(\sigma^2-1)^{\frac{3}{2}}}$. This is nothing, but the derivative with respect to the momenta, and therefore with respect to $\sigma$ \footnote{We use several time in the paper the formulae
\begin{align*}
&\frac{\partial}{\partial p_{1\mu}}f(\sigma_{12})=-\frac{1}{m_1m_2}p_2^{\mu}\frac{\partial}{\partial \sigma_{12}}f(\sigma_{12}),\qquad\frac{\partial}{\partial p_{2\mu}}f(\sigma_{12})=-\frac{1}{m_1m_2}p_1^{\mu}\frac{\partial}{\partial \sigma_{12}}f(\sigma_{12}),\\
&\frac{\partial}{\partial p_{3\mu}}f(\sigma_{34})=-\frac{1}{m_1m_2}p_4^{\mu}\frac{\partial}{\partial \sigma_{34}}f(\sigma_{34}),\qquad\frac{\partial}{\partial p_{4\mu}}f(\sigma_{34})=-\frac{1}{m_1m_2}p_3^{\mu}\frac{\partial}{\partial \sigma_{34}}f(\sigma_{34}).
\end{align*}}, of the function $\frac{2\sigma^2-1}{\sqrt{\sigma^2-1}}$ appearing in the tree-level eikonal $\delta^{(0)}$ given in equation \eqref{F9}. This suggests that the tensorial structures considered in this section contributing to the subleading and sub-subleading waveform could be obtained directly, and more easily, by acting with the soft operators $\hat{\mathcal{S}}^{\mu\nu}_{(a)}$ on the eikonal and that there should be an analogue of the soft theorem for the (classical) amplitudes in $b$ space. In the next section we develop this idea and introduce a classical version of the soft theorem that indeed involves the eikonal phase. 

\section{Eikonal soft theorem}
\label{eiksoft}
So far, we computed the soft expansion of the waveform out of the soft expansion of the 5-point amplitude in momentum space computed in $(q_{1\parallel}^*,q_{2\parallel}^*)$. In this section we present a new way to get the universal soft expansion of the classical waveform that passes through the eikonal phase introduced earlier in section \ref{eikonal}. Beside being computationally easier, the method we present here smoothly extends to the higher-loop case. We use the one-loop soft theorem in \eqref{F5} and apply it to the classical, resummed $\mathcal{M}_4$ written as
\begin{align}
\label{es1}
\mathcal{M}_4(Q)\hat{\delta}(2p_1\cdot Q)\hat{\delta}(2p_2\cdot Q)=i\int d^4b\hspace{0.1cm}e^{-ib\cdot Q}\bigg(1-e^{2i\delta(\sigma,b)}\bigg)\simeq-i\int d^4b\hspace{0.1cm}e^{-ib\cdot Q}e^{2i\delta(\sigma,b)},
\end{align}
where $Q$ is the classical momentum transfer and where in the last step in \eqref{es1} we ignored a term $\sim \hat{\delta}^{(4)}(Q)$, which is irrelevant for our analysis.

Notice however that, because of 5-point kinematics, $\mathcal{M}_5$ depends on the variables $Q_1=p_1+p_4$ and $Q_2=p_2+p_3$, fixed by $Q_1+Q_2=k$, and having associated dual variables $b_1$ and $b_2$. In the soft $\omega\rightarrow 0$ limit $Q_1\simeq -Q_2\equiv Q$ and the 5-point kinematics becomes equal to the 4-point one and in fact, $\mathcal{M}_4$ only depends on $Q$. In order to restore the dependence on $b_1$ and $b_2$ on $\mathcal{M}_5$ obtained through the soft theorem, we write the classical $\mathcal{M}_4$ in a symmetric form with respect to the momenta
\begin{align}
\label{ep1}
\mathcal{M}_4(Q_1,Q_2)\hat{\delta}(2p_1\cdot Q_1)\hat{\delta}(2p_2\cdot Q_2)\simeq -i\int d^4b \hspace{0.1cm}e^{-ib_1\cdot (p_1+p_4)-ib_2\cdot (p_2+p_4)}e^{i\delta(\sigma_{12},b)+i\delta(\sigma_{34},b)},
\end{align}
where we have used $b\cdot Q=b_1\cdot Q_1+b_2\cdot Q_2$ and $Q_1=p_1+p_4$ and $Q_2=p_3+p_4$ and $\delta(\sigma_{12},b)$+$\delta(\sigma_{34},b)$ with
\begingroup
\addtolength\jot{9pt} %
\begin{align}
\label{F14}
&\delta(\sigma_{12},b)=-G_Nm_2m_2\frac{2\sigma^2_{12}-1}{\sqrt{\sigma^2_{12}-1}}\log (\mu b)+\mathcal{O}(G_N^2),\\& \delta(\sigma_{34},b)=-G_Nm_2m_2\frac{2\sigma^2_{34}-1}{\sqrt{\sigma^2_{34}-1}}\log (\mu b)+\mathcal{O}(G_N^2),
\end{align}
\endgroup
which is the eikonal phase symmetrized with respect to incoming and outgoing momenta. We then construct the soft waveform as in \eqref{F5}, but this time acting on the resummed amplitude
\begin{align}
\label{es2}
\nonumber&\tilde{\mathcal{M}}^{\mu\nu}_5(p_i,p_f,b_i,k)=-i\kappa\int d^4 Q\hspace{0.1cm}e^{iQ\cdot b}\bigg\{e^{-\mathcal{W}_{\mathrm{phase}}}\bigg[\hat{\mathcal{S}}^{\mu\nu}_{(-1)}+\hat{\mathcal{S}}^{\mu\nu}_{(0)}+\hat{\mathcal{S}}^{\mu\nu}_{(1)}+(\hat{\mathcal{S}}_{(0)}^{\mu\nu}\mathcal{W}_{\mathrm{reg}})\\\nonumber&+\sum_{j=1}^4\frac{k_{\rho}k_{\sigma}}{p_j\cdot k}(\hat{J}_j^{\mu\rho}\mathcal{W}_{\mathrm{reg}})\hat{J}_j^{\nu\sigma}+\frac{1}{2}\sum_{j=1}^4\frac{k_{\rho}k_{\sigma}}{p_j\cdot k}(\hat{J}_j^{\mu\rho}\mathcal{W}_{\mathrm{reg}})(\hat{J}_j^{\nu\sigma}\mathcal{W}_{\mathrm{reg}})\\&+(\hat{\mathcal{S}}_{(1)}^{\mu\nu}\mathcal{W}_{\mathrm{reg}})\bigg]\int d^4b' \hspace{0.1cm}e^{-ib'_1\cdot (p_1+p_4)-ib'_2\cdot (p_2+p_4)}e^{i\delta(\sigma_{12},b')+i\delta(\sigma_{34},b')}\bigg\}+\mathcal{O}(\omega^2,\omega^2\log\omega).
\end{align}
This procedure, beside immediately reproducing the tree-level results obtained earlier, it smoothly extends to the one-loop case, which we present in detail here as the main result of the present paper. We also write explicitly the symmetrized function
 $\mathcal{W}_{\mathrm{reg}}$ in \eqref{W3}
 \begingroup
\addtolength\jot{9pt}
\begin{align}
\label{F15}
&\mathcal{W}_{\mathrm{reg}}\simeq-iG_Nm_1m_2\bigg(\frac{2\sigma^2_{12}-1}{\sqrt{\sigma^2_{12}-1}}\log(\omega_-L)+\frac{2\sigma_{34}^2-1}{\sqrt{\sigma_{34}^2-1}}\log(\omega_+L)\bigg),\\
&\mathcal{W}_{\mathrm{phase}}\simeq2iG_N (p_3+p_4)\cdot k\log (\omega_+R)=-2iG_N E\omega \log (\omega_+R),
\end{align}
\endgroup
where $\omega_{\pm}=\omega\pm i\epsilon$ and in the last step we used $E=E_1+E_2$ and the center of mass parametrization in \eqref{F15.1}, \eqref{F15.2} and \eqref{F15.3} and $k=\omega n$ with $n=(1,\sin\theta,\sin\phi,\sin\theta\cos\phi,\cos\theta)$.
We computed explicitly the right hand side in \eqref{es2} and it yields \footnote{Here we simplify the notation $\tilde{\mathcal{M}}^{\mu\nu}(p_i,p_f,b_i,k)\equiv \tilde{\mathcal{M}}^{\mu\nu}(p_i,p_f)$ since we are mainly interested in the perturbative expansion $p_f=p_i+\mathcal{O}(G_N)$.}
\begingroup
\addtolength\jot{9pt} %
\begin{align}
\label{F16}
\nonumber &\tilde{\mathcal{M}}_5^{\mu\nu}(p_i,p_f)=e^{2iG_N E\omega \log(\omega_+ R)}\big\{\mathcal{A}_{(-1)}^{\mu\nu}(p_i,p_f)+\mathcal{A}_{(0)}^{\mu\nu}(p_i,p_f)+\mathcal{A}_{(1)}^{\mu\nu}(p_i,p_f)\\\nonumber&+\mathcal{B}_{(0)}^{\mu\nu}(p_i)\log(\omega_-b)+\mathcal{B}_{(0)}^{\mu\nu}(p_f)\log(\omega_+b)+\mathcal{D}^{\mu\nu}_{(1)}(p_i)\log^2(\omega_-b)+\mathcal{D}_{(1)}^{\mu\nu}(p_f)\log^2(\omega_+b)\\\nonumber&+[\mathcal{B}^{\mu\nu}_{(1)}(p_i)+\mathcal{C}^{\mu\nu}_{(1)}(p_i)]\log(\omega_-b)+[\mathcal{B}^{\mu\nu}_{(1)}(p_f)+\mathcal{C}^{\mu\nu}_{(1)}(p_f)]\log(\omega_+b)\big\}e^{2i\delta(\sigma,b)}\\&+\mathcal{O}(\omega^2,\omega^2\log\omega).
\end{align}
\endgroup
The various tensorial structures $\mathcal{T}^{\mu\nu}_{(a)}$ and their PM expansions are derived and listed in appendix \ref{A}. The subscript $(a)$, with $a=-1,0,1$ indicates that  $\mathcal{T}^{\mu\nu}_{(a)}$ scales as $\sim\omega^a$. In principle, the above formulae are valid at all orders in $G_N$, depending on the PM order of the classical momentum exchanged $Q$ that is used when we substitute $p_4=-p_1+Q$ and $p_3=-p_2-Q$. We find that the structures $\mathcal{A}^{\mu\nu}_{(a)}$ that do not contain logarithms are equal to the ones found in the previous section once we use $p_4=-p_1+Q$ and $p_3=-p_2-Q$ with $Q=Q^{(0)}$
\begingroup
\addtolength\jot{9pt} %
\begin{align}
\label{Astr}
&\mathcal{A}_{(-1)}^{\mu\nu}(p_i,p_f)=\tilde{\mathcal{M}}^{(0)\mu\nu}_5|_{\mathcal{O}(\omega^{-1})}(p_i,b_i,k)+\mathcal{O}(\kappa G^2_N),\\&
\mathcal{A}_{(0)}^{\mu\nu}(p_i,p_f)=\tilde{\mathcal{M}}^{(0)\mu\nu}_5|_{\mathcal{O}(\omega^{0})}(p_i,b_i,k)+\mathcal{O}(\kappa G^2_N),\\&\mathcal{A}_{(1)}^{\mu\nu}(p_i,p_f)=\tilde{\mathcal{M}}^{(0)\mu\nu}_5|_{\mathcal{O}(\omega)}(p_i,b_i,k)+\mathcal{O}(\kappa G^2_N).
\end{align}
\endgroup
While in the previous section these structures were obtained by carefully keeping track and summing several different contributions, in this approach they are more naturally computed as the action of the soft operators on the symmetrized Fourier exponent
\begin{align}
\label{Astr1}
\mathcal{A}^{\mu\nu}_{(a)}(p_i,p_f)e^{-i(p_1+p_4)\cdot b_1-i(p_2+p_3)\cdot b_2}=-i\kappa\hspace{0.1cm}\hat{\mathcal{S}}^{\mu\nu}_{(a)} \bigg\{e^{-i(p_1+p_4)\cdot b_1-i(p_2+p_3)\cdot b_2}\bigg\}.
\end{align}
Let us stress an important point here. In principle, the action of the various terms in \eqref{es2} will give terms proportional to $\log(\mu b)$, coming from the leading eikonal and to $\log(\omega_{\pm}L)$, coming from $\mathcal{W}_{\mathrm{reg}}$. In this way, terms that are usually regarded as one-loop, combine in a simple way with tree-level ones. We find that the coefficients of these logarithmic terms are equal and therefore the argument of the various logarithms becomes dimensionless and equal to $\omega b$, as displayed in \eqref{F16}, and therefore the method of region is not needed within this approach. We also note that all the tensors appearing in that equation are multiplied by an eikonal phase $e^{2i\delta}$. Just as the Fourier transform of the elastic 4-point amplitude in the classical limit resums into an exponentiated structure, also does the 5-point one in the soft limit. Of course the exponentiation should be checked order by order in $G_N$ by explicit perturbative computations. However, soft theorems predict such exponentiation in the soft limit. In the remainder of this paper we will only consider terms that come from the first order of the exponential $e^{2i\delta}\simeq 1$. Higher order terms in $G_N$ will scale with smaller powers of $\hbar$ and are superclassical.

Before proceeding further let us note that
\begin{align}
\label{F17}
\log(\omega_{\pm}b)=\log(\sqrt{\omega^2+\epsilon^2}b)\pm i\arctan(\frac{\epsilon}{\omega})\sim\log(\omega b)\pm i\frac{\pi}{2},
\end{align}
and that \cite{Bini:2023fiz}
\begingroup
\addtolength\jot{6pt} 
\begin{align}
\label{F18}
\nonumber &e^{2iG_NE\omega\log(\omega_+ R)}=e^{2iG_NE\omega\log(\frac{R}{b})}e^{2iG_NE\omega\log(\omega b)}\\&=e^{2iG_NE\omega\log(\frac{R}{b})}(1+2iG_NE\omega\log(\omega_+ b)-4G^2_N E^2\omega^2\log^2(\omega_+ b)+\dots).
\end{align}
\endgroup
This allows to organize \eqref{F16} into universal terms proportional to $\log(\omega b)$, $\omega\log^2(\omega b)$ and $\omega\log(\omega b)$, in which we are interested. We have, including only these terms
\begingroup
\addtolength\jot{6pt} 
\begin{align}
\label{F19}
\nonumber\tilde{\mathcal{M}}^{\mu\nu}_5(p_i,p_f)\sim &\hspace{0.1cm}e^{2iG_NE\omega\log(\frac{R}{b})}\big[\tilde{\mathcal{M}}_{\log(\omega b)}^{\mu\nu}(p_i,p_f)\log(\omega b)\\&+\tilde{\mathcal{M}}_{\omega\log^2(\omega b)}^{\mu\nu}(p_i,p_f)\omega\log^2(\omega b)+\tilde{\mathcal{M}}_{\omega\log(\omega b)}^{\mu\nu}(p_i,p_f)\omega\log(\omega b)\big],
\end{align}
\endgroup
where
\begingroup
\addtolength\jot{9pt}  
\begin{align}
\label{F20}
&\tilde{\mathcal{M}}_{\log(\omega b)}^{\mu\nu}(p_i,p_f)=2iG_NE\omega\mathcal{A}^{\mu\nu}_{(-1)}(p_i,p_f)+\mathcal{B}_{(0)}^{\mu\nu}(p_i)+\mathcal{B}_{(0)}^{\mu\nu}(p_f),\\
\label{F21}
\nonumber &\tilde{\mathcal{M}}_{\omega\log^2(\omega b)}^{\mu\nu}(p_i,p_f)=-4G^2_NE^2\omega^2\mathcal{A}_{(-1)}^{\mu\nu}(p_i,p_f)+\mathcal{D}_{(1)}^{\mu\nu}(p_i)+\mathcal{D}_{(1)}^{\mu\nu}(p_f)\\&\hspace{3.5cm}+2iEG_N\omega[\mathcal{B}_{(0)}^{\mu\nu}(p_i)+\mathcal{B}_{(0)}^{\mu\nu}(p_f)],\\
\nonumber &\tilde{\mathcal{M}}_{\omega\log(\omega b)}^{\mu\nu}(p_i,p_f)=\mathcal{B}_{(1)}^{\mu\nu}(p_i)+\mathcal{B}_{(1)}^{\mu\nu}(p_f)+2iE G_N\omega\mathcal{A}_{(0)}^{\mu\nu}(p_i,p_f)\\\label{F22}&\hspace{3.5cm}+i\pi[\mathcal{D}_{(1)}^{\mu\nu}(p_f)-\mathcal{D}_{(1)}^{\mu\nu}(p_i)]-4i\pi G^2_N\omega^2\mathcal{A}_{-1}^{\mu\nu}(p_i,p_f).
\end{align}
\endgroup
 The $\omega\log(\omega b)$ term should also contain the tensorial structure $\mathcal{C}_{(1)}^{\mu\nu}$, however as clear from its explicit expression in \eqref{A10}, it is subleading in the classical limit. We stress again that the above formulae are valid to all orders in $G_N$ because they depend on the perturbative expansion of the exchanged classical momentum.

Notice also that when we change the IR scale $R\rightarrow \tilde{R}$ the $\omega\log(\omega b)$ term gets modified as
\begin{align}
\label{F23}
\tilde{\mathcal{M}}_{\omega\log(\omega b)}^{\mu\nu}(p_i,p_f)\longrightarrow\tilde{\mathcal{M}}_{\omega\log(\omega b)}^{\mu\nu}(p_i,p_f)+2iG_NE\omega\log(\frac{\tilde{R}}{R})\tilde{\mathcal{M}}_{\log(\omega b)}^{\mu\nu}(p_i,p_f).
\end{align}
Apart from  irrelevant overall factors, Eq. \eqref{F20} reproduces  the quantity $B_{\mu \nu} - C_{\mu \nu}$ in Eqs. (1.7) and (1.8) and \eqref{F21} reproduces the quantity $F_{\mu \nu} -G_{\mu \nu}$ in Eqs. (1.9) and (1.10) of Ref.~\cite{Sahoo:2021ctw}.

\section{1PM and 2PM universal amplitude}
\label{universal}
The Fourier transform of the 5-point amplitude admits a PM expansion
\begingroup
\addtolength\jot{8pt} 
\begin{align}
\label{F24}
\tilde{\mathcal{M}}^{\mu\nu}_5(p_i,p_f)=\tilde{\mathcal{M}}^{(0)\mu\nu}_5(p_i)+\tilde{\mathcal{M}}^{(1)\mu\nu}_5(p_i)+\mathcal{O}(\kappa G_N^{3}),
\end{align}
\endgroup
and, correspondingly, the universal terms of the waveform proportional to $\log(\omega b)$, $\omega\log^2(\omega b)$ and to $\omega\log(\omega b)$ that are displayed in equations \eqref{F20}, \eqref{F21} and \eqref{F22},
can be PM expanded as
\begin{align}
\tilde{\mathcal{M}}^{\mu\nu}_{\omega^x\log^y(\omega b)}(p_i,p_f)=\tilde{\mathcal{M}}^{(0)\mu\nu}_{\omega^x\log^y(\omega b)}(p_i)+\tilde{\mathcal{M}}^{(1)\mu\nu}_{\omega^x\log^y(\omega b)}(p_i)+\mathcal{O}(\kappa G_N^{3}),
\end{align}
with $x=0,1$ and $y=1,2$.
\subsection*{1PM}
Using the perturbative expansion shown in appendix \ref{A} we get at 1PM
\begingroup
\addtolength\jot{9pt} %
 \begin{align}
\label{F25}
&\tilde{\mathcal{M}}^{(0)\mu\nu}_{\log(\omega b)}(p_i)=2\kappa G_N\frac{\sigma(2\sigma^2-3)}{(\sigma^2-1)^{\frac{3}{2}}}\bigg(\frac{p_2\cdot k}{p_1\cdot k}p_1^{\mu}p_1^{\nu}+\frac{p_1\cdot k}{p_2\cdot k}p_2^{\mu}p_2^{\nu}-p_1^{\mu}p_2^{\nu}-p_1^{\nu}p_2^{\mu}\bigg),\\
&\tilde{\mathcal{M}}^{(0)\mu\nu}_{\omega\log^2(\omega b)}(p_i)=0,\\
&\nonumber\tilde{\mathcal{M}}_{\omega\log(\omega b)}^{(0)\mu\nu}(p_i)=-2i\kappa G_N\frac{\sigma(2\sigma^2-3)}{(\sigma^2-1)^{\frac{3}{2}}}\bigg[(b_1\cdot k)\frac{p_2\cdot k}{p_1\cdot k}p_1^{\mu}p_1^{\nu}+(b_2\cdot k)\frac{p_1\cdot k}{p_2\cdot k}p_2^{\mu}p_2^{\nu}\\&\label{F25.1}-(b_1\cdot k+b_2\cdot k)\frac{p_1^{\mu}p_2^{\nu}+p_1^{\nu}p_2^{\mu}}{2}-(p_2\cdot k)\frac{p_1^{\mu}b^{\nu}+p_1^{\nu}b^{\mu}
}{2}+(p_1\cdot k)\frac{p_2^{\mu}b^{\nu}+p_2^{\nu}b^{\mu}}{2}\bigg].
\end{align}
\endgroup
Equations \eqref{F25} and \eqref{F25.1} correctly reproduce the $\mathcal{O}(\log(\omega b))$ and $\mathcal{O}(\omega\log(\omega b))$ contributions to the leading waveform
\cite{DiVecchia:2023frv} and they match equations \eqref{sublog} and \eqref{subsublog}.
\subsection*{2PM}
At 2PM we get
\begingroup
\addtolength\jot{9pt}  
\begin{align}
\label{F26}
\nonumber &\tilde{\mathcal{M}}^{(1)\mu\nu}_{\log(\omega b)}(p_i)=2\kappa \frac{G^2_N m_1m_2}{b^2} (p_1\cdot k+p_2\cdot k)\bigg(\frac{2\sigma^2-1}{\sqrt{\sigma^2-1}}\bigg)\bigg[(b\cdot k)\bigg(\frac{p_1^{\mu}p_1^{\nu}}{(p_1\cdot k)^2}-\frac{p_2^{\mu}p_2^{\nu}}{(p_2\cdot k)^2}\bigg)\\&-\frac{p_1^{\mu}b^{\nu}+p_1^{\nu}b^{\mu}}{p_1\cdot k}+\frac{p_2^{\mu}b^{\nu}+p_2^{\nu}b^{\mu}}{p_2\cdot k}\bigg]\bigg[2-\frac{\sigma(2\sigma^2-3)}{(\sigma^2-1)^{\frac{3}{2}}}\bigg],\\
\nonumber  
&\tilde{\mathcal{M}}^{(1)\mu\nu}_{\omega\log^2(\omega b)}(p_i)=4i\kappa G^2_N(p_1\cdot k+p_2\cdot k)\frac{\sigma(2\sigma^2-3)}{(\sigma^2-1)^{\frac{3}{2}}}\bigg[\frac{p_2\cdot k}{p_1\cdot k}p_1^{\mu}p_1^{\mu}+\frac{p_1\cdot k}{p_2\cdot k}p_2^{\mu}p_2^{\mu}\\\label{F26.a}&-p_1^{\mu}p_2^{\nu}-p_1^{\nu}p_2^{\mu}\bigg],\\
&\tilde{\mathcal{M}}^{(1)\mu\nu}_{\omega\log(\omega b)}(p_i)={}_{I}\tilde{\mathcal{M}}^{(1)\mu\nu}_{\omega\log(\omega b)}(p_i)+{}_{II}\tilde{\mathcal{M}}^{(1)\mu\nu}_{\omega\log(\omega b)}(p_i)+{}_{III}\tilde{\mathcal{M}}^{(1)\mu\nu}_{\omega\log(\omega b)}(p_i),
\end{align}
\endgroup
with
\begingroup
\addtolength\jot{9pt} %
\begin{align}
\nonumber\label{F26.c}
&{}_{I}\tilde{\mathcal{M}}^{(1)\mu\nu}_{\omega\log(\omega b)}(p_i)=2i\kappa \frac{G^2_N m_1m_2}{b^2}\frac{\sigma(2\sigma^2-3)}{(\sigma^2-1)^{\frac{3}{2}}}\frac{2\sigma^2-1}{\sqrt{\sigma^2-1}}(p_1\cdot k+p_2\cdot k)\bigg\{(b\cdot k) \bigg[(b_1\cdot k)\\ &\times\frac{p_1^{\mu}p_1^{\nu}}{(p_1\cdot k)^2}-(b_2\cdot k)\frac{p_2^{\mu}p_2^{\nu}}{(p_2\cdot k)^2}\bigg]-(b_1\cdot k)\frac{p_1^{\mu}b^{\nu}+p_1^{\nu}b^{\mu}}{p_1\cdot k}+(b_2\cdot k)\frac{p_2^{\mu}b^{\nu}+p_2^{\nu}b^{\mu}}{p_2\cdot k}+b^{\mu}b^{\nu}\bigg\},\\\nonumber
&\label{F26d}{}_{II}\tilde{\mathcal{M}}^{(1)\mu\nu}_{\omega\log(\omega b)}(p_i)=-4i\kappa \frac{G^2_N m_1m_2}{b^2}\frac{2\sigma^2-1}{\sqrt{\sigma^2-1}}(p_1\cdot k+p_2\cdot k)\bigg\{(b\cdot k) \bigg[(b_1\cdot k)\frac{p_1^{\mu}p_1^{\nu}}{(p_1\cdot k)^2}\\  &-(b_2\cdot k)\frac{p_2^{\mu}p_2^{\nu}}{(p_2\cdot k)^2}\bigg]-(b_1\cdot k)\frac{p_1^{\mu}b^{\nu}+p_1^{\nu}b^{\mu}}{p_1\cdot k}+(b_2\cdot k)\frac{p_2^{\mu}b^{\nu}+p_2^{\nu}b^{\mu}}{p_2\cdot k}+b^{\mu}b^{\nu}\bigg\},\\ \nonumber &{}_{III}\tilde{\mathcal{M}}^{(1)\mu\nu}_{\omega\log(\omega b)}(p_i)=\pi\kappa G^2_N\bigg[\frac{\sigma(2\sigma^2-3)}{(\sigma^2-1)^{\frac{3}{2}}}\bigg]^2(p_1\cdot k+p_2\cdot k)\bigg(\frac{p_2\cdot k}{p_1\cdot k}p_1^{\mu}p_1^{\nu}+\frac{p_1\cdot k}{p_2\cdot k}p_2^{\mu}p_2^{\nu}\\&-p_1^{\mu}p_2^{\nu}-p_1^{\nu}p_2^{\mu}\bigg).
\end{align}
\endgroup
Equations \eqref{F26} and \eqref{F26.a} reproduce the $\mathcal{O}(\log(\omega b))$ and $\mathcal{O}(\omega\log^2(\omega b))$ contributions to the subleading waveform. In particular, we find that \eqref{F26} and \eqref{F26.a} are compatible with results in equation (9.9) of \cite{Bini:2023fiz} and with (2.6) of \cite{Sahoo:2018lxl} and with (2.5) of \cite{Ghosh:2021bam}. Notice that \eqref{F26} is apparently not compatible with what is stated in equation (4.11) of \cite{Georgoudis:2023eke}. However, when expressing the initial momenta $p_1$ and $p_2$ in terms of the variables $\tilde p_1$ and $\tilde p_2$ introduced in (2.30) of \cite{Georgoudis:2023eke} the sum of the 1PM contribution in \eqref{F25} and the 2PM one in \eqref{F26} is actually equal to (4.11) of \cite{Georgoudis:2023eke}\footnote{We thank the authors of \cite{Georgoudis:2023eke} for pointing this out.}. We notice that the $\mathcal{O}(\omega\log^2\omega)$ term satisfies 
\begingroup
\addtolength\jot{9pt}
\begin{align}
\tilde{\mathcal{M}}^{(1)\mu\nu}_{\omega\log^2(\omega b)}(p_i)=2i G_N(p_1\cdot k+p_2\cdot k)\tilde{\mathcal{M}}^{(0)\mu\nu}_{\log(\omega b)}(p_i)+\mathcal{O}(G^3_N),\end{align}
\endgroup
in agreement also with (4.12) of \cite{Georgoudis:2023eke}.

Notice that equation \eqref{F26.c} and \eqref{F26d} can be put together, yielding
\begin{align}
\nonumber
&{}_{I}\tilde{\mathcal{M}}^{(1)\mu\nu}_{\omega\log(\omega b)}(p_i)+{}_{II}\tilde{\mathcal{M}}^{(1)\mu\nu}_{\omega\log(\omega b)}(p_i)=-2i\kappa \frac{G^2_N m_1m_2}{b^2}\frac{2\sigma^2-1}{\sqrt{\sigma^2-1}}(p_1\cdot k+p_2\cdot k)\bigg\{(b\cdot k) \bigg[(b_1\cdot k)\\ \nonumber &\times\frac{p_1^{\mu}p_1^{\nu}}{(p_1\cdot k)^2}-(b_2\cdot k)\frac{p_2^{\mu}p_2^{\nu}}{(p_2\cdot k)^2}\bigg]-(b_1\cdot k)\frac{p_1^{\mu}b^{\nu}+p_1^{\nu}b^{\mu}}{p_1\cdot k}+(b_2\cdot k)\frac{p_2^{\mu}b^{\nu}+p_2^{\nu}b^{\mu}}{p_2\cdot k}+b^{\mu}b^{\nu}\bigg\}\\&\label{2PMsubsub}\times\bigg[2-\frac{\sigma(2\sigma^2-3)}{(\sigma^2-1)^{\frac{3}{2}}}\bigg].
\end{align}
As we discuss later,
${}_{III}\tilde{\mathcal{M}}^{(1)\mu\nu}_{\omega\log(\omega b)}$ is the expression resummed  in the velocities of the first equation in (9.11) of \cite{Bini:2023fiz}. It is also contained in equation (4.14) of \cite{Georgoudis:2023eke}. Interestingly, the last factor between square bracket in \eqref{2PMsubsub} is the same that appears in $\mathcal{M}^{(1)\mu\nu}_{\log(\omega b)}$ in \eqref{F26}. As for the latter, we have checked that the sum of \eqref{F25.1} and \eqref{2PMsubsub} is compatible with what is stated in equations  (4.15) and (4.16) of \cite{Georgoudis:2023eke}, once we express it in terms of the variables $\tilde{p_i}$ and $b_e$ defined in (2.30) and (2.31) of the same reference.  The terms in $\tilde{\mathcal{M}}^{\mu\nu}_{\log(\omega b)}$ and ${}_{I}\tilde{\mathcal{M}}^{\mu\nu}_{\omega\log(\omega b)}$, having one factor of $\frac{\sigma(2\sigma^2-3)}{(\sigma^2-1)^{\frac{3}{2}}}$, come from the 2PM expansions of the tensors $\mathcal{B}^{\mu\nu}_{(a)}(p_i)+\mathcal{B}^{\mu\nu}_{(a)}(p_f)$, with $a=0,1$ whose 1PM contribution ($p_f=p_i$) exactly reproduces the tree-level waveform. These terms are entirely predicted by the subleading and sub-subleading graviton soft factors $\hat{\mathcal{S}}^{\mu\nu}_{(a)}$ once we perform the PM expansion $p_4=-p_1+Q^{(1)}$ and $p_3=-p_2-Q^{(1)}$.
\section{PN expansion}
\label{PN}
The metric fluctuation can be obtained from the Fourier transform of the 5-point amplitude  as
\begin{align}
\mathcal{W}^{\mu\nu}=\frac{2}{R}\frac{G_N}{\kappa}\tilde{\mathcal{M}}^{\mu\nu}_5=\mathcal{W}^{(0)\mu\nu}+\mathcal{W}^{(1)\mu\nu}+\mathcal{O}(G^4_N).
\end{align}
Here, following \cite{Bini:2023fiz}, we consider the contraction of the waveform with polarization vectors
\begin{align}
\label{F27}
\mathcal{W}=\mathcal{W}^{\mu\nu}m_{\mu}m_{\nu},\qquad m^{\mu}=\frac{1}{\sqrt{2}}(\epsilon^{\mu}_{\theta}-i\epsilon^{\mu}_{\phi}),
\end{align}
with $\epsilon_\theta=\partial_\theta n$, $\epsilon_\phi=\partial_\phi n$  and $n=(1,\sin\theta,\sin\phi,\sin\theta\cos\phi,\cos\theta)$. Explicitly
\begin{align}
m^{\mu}=\frac{1}{\sqrt{2}}(0,\cos\theta\cos\phi+i\sin\phi,\cos\theta\sin\phi-i\cos\theta,-\sin\theta),
\end{align}
so that $m\cdot k=0$. In the previous section we have given expressions for the logarithmic universal terms that hold for relativistic velocities. Here, in order to compare with existing literature, we perform a PN expansion of the waveform in powers of $p_{\infty}$, related to $v$ by
\begin{align}
    v=\frac{p_{\infty}}{\sqrt{1+p_{\infty}^2}}.
\end{align}
We use a frame defined by
\begin{align}
b_1^{\mu}=\frac{E_2}{E}(0,b,0,0),\qquad b_2^{\mu}=\frac{E_1}{E}(0,-b,0,0),\qquad b^{\mu}=b_1^{\mu}-b_2^{\mu}=(0,b,0,0).
\end{align}
Expanding the tree-level waveform gives
\begingroup
\addtolength\jot{9pt}
\begin{align}
\nonumber&\mathcal{W}^{(0)}_{\log(\omega b)}=-i\frac{2G_N^2M^2\nu}{R}(\cos \phi +i \cos \theta  \sin \phi )^2\bigg[\frac{1}{p_{\infty}}-\Delta\sin\theta\sin\phi+\frac{p_{\infty}}{4}[ \left(2 (3 \nu -1) \sin ^2\theta\right.\\&\nonumber\left.\times  \cos 2 \phi +(3 \nu -1) \cos 2 \theta +\nu -7\right)]+\frac{\Delta p_{\infty}^2}{2}[2 (2 \nu -1) \sin ^2\theta  \sin ^2\phi -2 \nu +5]\sin\theta\sin\phi\\&+\mathcal{O}(p_{\infty}^3)\bigg],\\
\nonumber&\mathcal{W}^{(0)}_{\omega\log(\omega b)}=-\frac{2G^2_NbM^2\nu}{R}\bigg[-\frac{1}{p_{\infty}^2}(\cos \theta  \cos \phi +i \sin \phi ) (\cos \theta  \sin \phi -i \cos \phi) \\\nonumber&-\frac{\Delta}{p_{\infty}}\sin \theta \cos \phi  (\cos \phi +i \cos \theta  \sin \phi )^2+\frac{1}{2}(\cos \theta  \sin \phi -i \cos \phi ) \left[3 \cos \theta  \cos \phi \right.\\&+\sin \phi(3i+(3\nu-1)(\cos\theta\sin2\phi-2i\cos\phi^2)\sin^2\theta)]+\mathcal{O}(p_{\infty})\bigg].
\end{align}\endgroup
For the subleading waveform we get
\begingroup
\addtolength\jot{9pt}
\begin{align}
\nonumber&\mathcal{W}^{(1)}_{\log(\omega b)}=-\frac{2G^3_NM^3\nu}{bR}\bigg[\frac{2}{p_{\infty}^3}(\cos \theta  \cos \phi +i \sin \phi ) (\cos \phi +i \cos \theta  \sin \phi )+\frac{\Delta}{2p_{\infty}^2}\sin\theta\\\nonumber&\times[4 \cos \theta  \sin ^3\phi +2 i \cos ^3\phi -8 \cos \theta  \sin \phi  \cos ^2\phi -i (3 \cos 2 \theta+7) \sin ^2\phi  \cos \phi ]\\\nonumber&+\frac{1}{2p_{\infty}} (\cos \phi +i \cos \theta  \sin \phi )[\cos \theta \left(\cos \phi  ((3 \nu -1) \cos 2 \theta +3 \nu +1)+2 (3 \nu -1) \sin ^2\theta  \cos 3 \phi \right)\\&+i \left(2 (3 \nu -1) \sin ^2\theta  \sin 3 \phi +\sin \phi  ((3 \nu -1) \cos 2 \theta +3 \nu +1)\right)]+\mathcal{O}(p_{\infty}^0)\bigg],\\
\nonumber&\mathcal{W}^{(1)}_{\omega\log^2(\omega b)}=-\frac{2G^3_NM^3\nu}{R}(\cos\phi+i\cos\theta\sin\phi)^2\bigg[-\frac{2}{p_{\infty}}+2\Delta\sin\theta\sin\phi-\frac{p_{\infty}}{2}[2 (3 \nu -1) \\&\times\sin ^2\theta  \cos 2 \phi +(3 \nu -1) \cos 2 \theta +3 \nu -7]+\mathcal{O}(p_{\infty}^2)\bigg],
\end{align}
\endgroup
for the universal terms. For the non-universal ones we find
\begingroup
\addtolength\jot{9pt}
\begin{align}
\nonumber&{}_{I}\mathcal{W}^{(1)}_{\omega\log(\omega b)}=-i\frac{2 G^3_NM^3\nu}{R} \bigg[\frac{1}{p_{\infty}^4}(\cos\theta\cos\phi+i\sin\phi)^2+\frac{2i\Delta}{p_{\infty}^3}\cos\phi\sin\theta(\cos\theta\cos\phi+i\sin\phi)\\\nonumber&\times(\cos\phi+i\cos\theta\sin\phi)+\frac{1}{16p_{\infty}^2}[2(1+9\nu+(3-5\nu)\cos2\theta)\cos2\phi-4i\cos\theta(-1-5\nu\\&\nonumber+(3\nu-1)\cos2\theta)\sin 2\phi+\sin\theta^2(-3-7\nu+9(3\nu-1)\cos 4\phi+6(1-3\nu)\cos2\theta\sin^22\phi)\\\nonumber&+6i(3\nu-1)\sin\theta\sin2\theta\sin4\phi]-\frac{\Delta}{p_{\infty}}\cos\phi\sin\theta(\cos\phi+i\cos\theta\sin\phi)[(\nu\\&+2(2\nu-1)\cos2\phi\sin^2\theta)\sin\phi-i\cos\theta\cos\phi(\nu+4(1-2\nu)\sin^2\theta\sin^2\phi)]+\mathcal{O}(p_{\infty}^0)\bigg],
\\&\nonumber{}_{II}\mathcal{W}^{(1)}_{\omega\log(\omega b)}=-i\frac{2G^3_NM^3\nu}{R}\bigg[\frac{2}{p_{\infty}}(\cos \theta  \cos \phi +i \sin \phi )^2+4i\Delta\sin \theta  \cos \phi  (\cos \theta  \cos \phi +i \sin \phi )\\&\nonumber\times (\cos \phi +i \cos \theta  \sin \phi )+\frac{p_{\infty}}{8}[2\cos2\phi(10+9\nu+(6-5\nu)\cos2\theta)-4i\cos\theta\sin2\phi(\\\nonumber&(3\nu-1)\cos2\theta-7-5\nu)+\sin ^2\theta  \left(6 (1-3 \nu ) \cos 2 \theta  \sin ^22 \phi +9 (3 \nu -1) \cos 4 \phi -7 \nu \right.\\&\left.-15\right)+6 i (3 \nu -1) \sin \theta  \sin 2 \theta \sin 4 \phi ]+\mathcal{O}(p_{\infty}^2)\bigg],
\end{align}
\begin{align}
\label{term}&\nonumber{}_{III}\mathcal{W}^{(1)}_{\omega\log(\omega b)}=\pi\frac{2G^3_NM^3\nu}{R}(\cos \phi +i \cos \theta  \sin \phi )^2\bigg[ -  \frac{1}{2p_{\infty}^4}+\frac{\Delta}{2p_{\infty}^3}\sin \theta  \sin \phi  \\\nonumber&-\frac{1}{16p_{\infty}^2}[4 (3 \nu -1) \sin ^2\theta \cos 2 \phi+(6 \nu -2) \cos 2 \theta +6 \nu -26]-\frac{\Delta}{4p_{\infty}}[\sin \theta  \sin \phi  \left(2 (2 \nu -1)\right. \\&\nonumber\times\sin ^2\theta  \sin ^2\phi -3 \nu +8)-\frac{1}{16}[8 (5 (\nu -1) \nu +1) \sin ^4\theta  \sin ^4\phi-4(9+\nu(9\nu-31)\sin^2\theta\sin^2\phi\\&+\nu  (3 \nu -47)+15]+\mathcal{O}(p_{\infty})\bigg],
\end{align}
where $\nu=m_1m_2/M^2$ is the reduced mass, $M=m_1+m_2$ is the total mass and $\Delta=(m_1-m_2)/M$ the mass difference. We notice that \eqref{term}, which eventually comes from the argument in \eqref{F17},  
matches the first equation of (9.11) of \cite{Bini:2023fiz}. On the other hand, choosing appropriately the factor $\log(\frac{\tilde{R}}{R})$ in \eqref{F23} we could reproduce the terms in the PN expansion displayed in the last two equations in (9.11) of \cite{Bini:2023fiz} proportional to $\gamma_E$ and $\pi$.

\endgroup
\section{Conclusions}
\label{conclu}

The tree-level soft graviton theorem follows from gauge invariance and  fixes  the soft graviton behaviour  of the  amplitude (in momentum space), involving any number of  massive scalar particles and a graviton, in terms of the amplitude without the graviton and of a universal soft factor. 
We have applied  it to the case of the scattering of two massive particles with the emission of a graviton and we have then shown that, going to impact parameter space, one must regularise the result either with dimensional regularisation or with an infrared cutoff that, when taken to be equal to the graviton energy $\omega$, generates   $\mathcal{O}(\log \omega)$ and $\mathcal{O}(\omega \log \omega)$ universal terms.  

We have shown that the previous calculation can be significantly simplified if we apply the soft factor to the  eikonal resummed four-point amplitude instead of the tree-leve one. We use this framework to discuss also the one-loop corrected classical soft graviton behaviour that involves the two functions ${\cal{W}}_{\mathrm{reg}}$ and ${\cal{W}}_{\mathrm{phase}}$ that appear in various approches~\cite{Bern:2014oka,Laddha:2018vbn,Sahoo:2018lxl,Saha:2019tub,Sahoo:2021ctw,Ghosh:2021bam,Krishna:2023fxg,Agrawal:2023zea} to this problem, starting from the original paper by Weinberg~\cite{Weinberg:1965nx}. In this way, we compute both the 1PM and 2PM $\mathcal{O}(\log\omega)$, $\mathcal{O}(\omega\log^2\omega)$  and $\mathcal{O}(\omega\log\omega)$ contributions to the waveform. The 2PM sub-subleading term of order $\mathcal{O}(\omega\log\omega)$ is not universal and its not entirely captured by the soft graviton theorem. Here we display explicitly its universal part.
We have finally compared our results with those present in the literature.
In particular, we find perfect agreement with Refs.~\cite{Bern:2014oka,Laddha:2018vbn,Sahoo:2018lxl,Saha:2019tub,Sahoo:2021ctw,Ghosh:2021bam,Krishna:2023fxg,Agrawal:2023zea}. After having mapped the initial momenta $p_i$ used in our analysis to the ones $\tilde p_i$ used in \cite{Georgoudis:2023eke}, we find agreement also with this last reference. More precisely we get a universal term that agrees with the one obtained in Ref. \cite{Georgoudis:2023eke}, but, with our method, we do not get extra non-universal terms present in \cite{Georgoudis:2023eke}, for which it seems that the full one-loop computation needs to be performed.

\subsection*{Acknowledgements}
We thank Carlo Heissenberg,  Raffaele Marotta, Rodolfo Russo and Gabriele Veneziano for many useful discussions and  Laura Donnay, Kevin Nguyen, Romain Ruzziconi and  Biswajit Sahoo for useful correspondence.
The research of FA (PDV) is fully (partially) supported by the Knut and Alice Wallenberg Foundation under grant KAW 2018.0116. 

\appendix
\section{Action of soft operators}
\label{A}
In the following we define $\mathcal{M}_4(Q_1,Q_2)\hat{\delta}(2p_1\cdot Q_1)\hat{\delta}(2p_2\cdot Q_2)\equiv \mathcal{M}_4(p_i,p_f)$. Then, the action of the soft operators $\hat{\mathcal{S}}^{\mu\nu}_{(a)}$ in \eqref{F2},\eqref{F2.1},\eqref{F2.2} in \eqref{es2} is
\begin{align} 
\label{A1}
&\kappa\hat{\mathcal{S}}_{(-1)}^{\mu\nu}\mathcal{M}_4(p_i,p_f)=\int d^4b \hspace{0.05cm}e^{-iQ_1\cdot b_1-iQ_2\cdot b_2}e^{i\delta(p_i,b)+i\delta(p_f,b)}\mathcal{A}^{\mu\nu}_{(-1)}(p_i,p_f),\\
 \nonumber&\kappa\hat{\mathcal{S}}_{(0)}^{\mu\nu}\mathcal{M}_4(p_i,p_f)=\int d^4b\hspace{0.05cm}e^{-iQ_1\cdot b_1-iQ_2\cdot b_2}e^{i\delta(p_i,b)+i\delta(p_f,b)}\{\mathcal{A}^{\mu\nu}_{(0)}(p_i,p_f)+[\mathcal{B}^{\mu\nu}_{(0)}(p_i)\\&+\mathcal{B}^{\mu\nu}_{(0)}(p_f)]\log b\},\\\nonumber 
&\kappa\hat{\mathcal{S}}_{(1)}^{\mu\nu}\mathcal{M}_4(p_i,p_f)=\int d^4b\hspace{0.05cm}e^{-iQ_1\cdot b_1-iQ_2\cdot b_2}e^{i\delta(p_i,b)+i\delta(p_f,b)}\{\mathcal{A}^{\mu\nu}_{(1)}(p_i,p_f)+[\mathcal{B}^{\mu\nu}_{(1)}(p_i)\\&+\mathcal{B}^{\mu\nu}_{(1)}(p_f)+\mathcal{C}^{\mu\nu}_{(1)}(p_i)+\mathcal{C}^{\mu\nu}_{(1)}(p_f)]\log b+[\mathcal{D}^{\mu\nu}_{(1)}(p_i)+\mathcal{D}^{\mu\nu}_{(1)}(p_f)]\log^2 b\},
\end{align}
and that of the additional terms in the one-loop soft theorem in \eqref{F5} involving $\mathcal{W}_{\mathrm{reg}}$ is
\begin{align}
\label{A2}
&\nonumber\kappa\sum_{j=1}^4\frac{p_j^{(\mu}k_{\rho}}{p_j\cdot k}(\hat{J}_j^{\nu)\rho}\mathcal{W}_{\mathrm{reg}})\mathcal{M}_4(p_i,p_f)=\int d^4b \hspace{0.05cm}e^{-iQ_1\cdot b_1-iQ_2\cdot b_2}e^{i\delta(p_i,b)+i\delta(p_f,b)}\{\\&\mathcal{B}_{(0)}^{\mu\nu}(p_i)\log\omega_-+\mathcal{B}_{(0)}^{\mu\nu}(p_f)\log\omega_+\},\\&\nonumber \kappa\sum_{j=1}^4\frac{k_{\rho}k_{\sigma}}{p_j\cdot k}(\hat{J}_j^{(\mu\rho}\mathcal{W}_{\mathrm{reg}})\hat{J}_j^{\nu)\sigma}\mathcal{M}_4(p_i,p_f)=\int d^4b\hspace{0.05cm}e^{-iQ_1\cdot b_1-iQ_2\cdot b_2}e^{i\delta(p_i,b)+i\delta(p_f,b)}\{\\&\mathcal{B}_{(1)}^{\mu\nu}(p_i)\log\omega_-+\mathcal{B}_{(1)}^{\mu\nu}(p_f)\log\omega_++2[\mathcal{D}_{(1)}^{\mu\nu}(p_i)\log \omega_-+\mathcal{D}_{(1)}^{\mu\nu}(p_f)\log\omega_+]\log b\},\\\nonumber
&\frac{\kappa}{2}\sum_{j=1}^4\frac{k_{\rho}k_{\sigma}}{p_j\cdot k}(\hat{J}_j^{\mu\rho}\mathcal{W}_{\mathrm{reg}})(\hat{J}_j^{\nu\sigma}\mathcal{W}_{\mathrm{reg}})\mathcal{M}_4(p_i,p_f)=\int d^4b\hspace{0.05cm} e^{-iQ_1\cdot b_1-iQ_2\cdot b_2}e^{i\delta(p_i,b)+i\delta(p_f,b)}\{\\&\mathcal{D}_{(1)}^{\mu\nu}(p_i)\log^2 \omega_-+\mathcal{D}_{(1)}^{\mu\nu}(p_f)\log^2\omega_+\},\\\nonumber&\frac{\kappa}{2}\sum_{j=1}^4\frac{k_{\rho}k_{\sigma}}{p_j\cdot k}(\hat{J}_j^{\mu\rho}\hat{J}_j^{\nu\sigma}\mathcal{W}_{\mathrm{reg}})\mathcal{M}_4(p_i,p_f)=\int d^4b \hspace{0.05cm}e^{-iQ_1\cdot b_1-iQ_2\cdot b_2}e^{i\delta(p_i,b)+i\delta(p_f,b)}\{\\&\mathcal{C}^{\mu\nu}_{(1)}(p_i)\log\omega_-+\mathcal{C}^{\mu\nu}_{(1)}(p_f)\log\omega_+\}.
\end{align}
In the above computation we dropped the irrelevant scales $\mu$ and $L$ from the arguments of the logarithms. When summing the above contributions we see that each term proportional to $\log b$ and to $\log\omega_{\pm}$ can be put together to yield $\log(\omega_{\pm}b)$. Each tensorial structure $\mathcal{T}^{\mu\nu}_{(a)}$ scales as $\omega^{a}$. The ones that do not contain logarithmic terms are given by
\begin{align}
\label{A3.1}
&\mathcal{A}_{(-1)}^{\mu\nu}(p_i,p_f)=-i\kappa\sum_{j=1}^4\frac{p_i^{\mu}p_i^{\nu}}{p_i\cdot k},\\
\nonumber&\mathcal{A}_{(0)}^{\mu\nu}(p_i,p_f)=-\kappa\bigg[(b_1\cdot k)\bigg(\frac{p_1^{\mu}p_1^{\nu}}{p_1\cdot k}+\frac{p_4^{\mu}p_4^{\nu}}{p_4\cdot k}\bigg)+(b_2\cdot k)\bigg(\frac{p_2^{\mu}p_2^{\nu}}{p_2\cdot k}+\frac{p_3^{\mu}p_3^{\nu}}{p_3\cdot k}\bigg)\\&-\frac{p_1^{\mu}b_1^{\nu}+p_1^{\nu}b_1^{\mu}}{2}-\frac{p_4^{\mu}b_1^{\nu}+p_4^{\nu}b_1^{\mu}}{2}-\frac{p_2^{\mu}b_2^{\nu}+p_2^{\nu}b_2^{\mu}}{2}-\frac{p_3^{\mu}b_2^{\nu}+p_3^{\nu}b_2^{\mu}}{2}\bigg],\\
\nonumber &\mathcal{A}_{(1)}^{\mu\nu}(p_i,p_f)=i\frac{\kappa}{2}\bigg[(b_1\cdot k)^2\bigg(\frac{p_1^{\mu}p_1^{\nu}}{p_1\cdot k}+\frac{p_4^{\mu}p_4^{\nu}}{p_4\cdot k}\bigg)+(b_2\cdot k)^2\bigg(\frac{p_2^{\mu}p_2^{\nu}}{p_2\cdot k}+\frac{p_3^{\mu}p_3^{\nu}}{p_3\cdot k}\bigg)\\ &-(b_1\cdot k)(p_1^{\mu}b_1^{\nu}+p_1^{\nu}b_1^{\mu}+p_4^{\mu}b_1^{\nu}+p_4^{\nu}b_1^{\mu})-(b_2\cdot k)(p_2^{\mu}b_2^{\nu}+p_2^{\nu}b_2^{\mu}+p_3^{\mu}b_2^{\nu}+p_3^{\nu}b_2^{\mu})\bigg].
\end{align}
Using momentum conservation as $p_4=-p_1+Q$ and $p_3=-p_2-Q$, with 
\begin{align}
\label{A4}
Q^{\mu}=-2G_Nm_1m_2\frac{2\sigma^2-1}{\sqrt{\sigma^2-1}}\frac{b^{\mu}}{b^2}+\mathcal{O}(G_N^2),
\end{align}
we get at linear order in $G_N$
\begin{align}
\label{A5}
\nonumber&\mathcal{A}_{(-1)}^{\mu\nu}(p_i)=-2i\kappa \frac{G_Nm_1m_2}{b^2}\frac{2\sigma^2-1}{\sqrt{\sigma^2-1}}\bigg[(b\cdot k)\bigg(\frac{p_1^{\mu}p_1^{\nu}}{(p_1\cdot k)^2}-\frac{p_2^{\mu}p_2^{\nu}}{(p_2\cdot k)^2}\bigg)-\frac{p_1^{\mu}b^{\nu}+p_1^{\nu}b^{\mu}}{p_1\cdot k}\\&+\frac{p_2^{\mu}b^{\nu}+p_2^{\nu}b^{\mu}}{p_2\cdot k}\bigg]+\mathcal{O}(G_N^2),\\\nonumber &\mathcal{A}_{(0)}^{\mu\nu}(p_i)=-2\kappa \frac{G_Nm_1m_2}{b^2}\frac{2\sigma^2-1}{\sqrt{\sigma^2-1}}\bigg\{(b\cdot k)\bigg[(b_1\cdot k)\frac{p_1^{\mu}p_1^{\nu}}{(p_1\cdot k)^2}-(b_2\cdot k)\frac{p_2^{\mu}p_2^{\nu}}{(p_2\cdot k)^2}\bigg]\\&-(b_1\cdot k)\frac{p_1^{\mu}b^{\nu}+p_1^{\nu}b^{\mu}}{p_1\cdot k}+(b_2\cdot k)\frac{p_2^{\mu}b^{\nu}+p_2^{\nu}b^{\mu}}{p_2\cdot k}+b^{\mu}b^{\nu}\bigg\}+\mathcal{O}(G^2_N),\\\nonumber &\mathcal{A}_{(1)}^{\mu\nu}(p_i)=i\kappa \frac{G_N m_1m_2}{b^2}\frac{2\sigma^2-1}{\sqrt{\sigma^2-1}}\bigg\{(b\cdot k)\bigg[(b_1\cdot k)^2\frac{p_1^{\mu}p_1^{\nu}}{(p_1\cdot k)^2}-(b_2\cdot k)^2\frac{p_2^{\mu}p_2^{\nu}}{(p_2\cdot k)^2}\bigg]\\ &-(b_1\cdot k)^2\frac{p_1^{\mu}b^{\nu}+p_1^{\nu}b^{\mu}}{p_1\cdot k}+(b_2\cdot k)^2\frac{p_2^{\mu}b^{\nu}+p_2^{\nu}b^{\mu}}{p_2\cdot k}+(b_1\cdot k+b_2\cdot k)b^{\mu}b^{\nu}\bigg\}+\mathcal{O}(G^2_N).
\end{align}
Notice that, after having imposed momentum conservation, the previous tensors satisfy gauge invariance $k_{\mu}\mathcal{A}^{\mu\nu}_{(a)}=k_{\nu}\mathcal{A}^{\mu\nu}_{(a)}=0$. Furthermore, notice that they scale as $\kappa G_N$, which is the correct coupling for a 5-point amplitude at tree-level.

The terms comprising $\log \omega$ are
\begin{align}
\label{A6}
&\mathcal{B}_{(0)}^{\mu\nu}(p_i)=\kappa G_N\frac{\sigma(2\sigma^2-3)}{(\sigma^2-1)^{\frac{3}{2}}}\bigg(\frac{p_2\cdot k}{p_1\cdot k}p_1^{\mu}p_1^{\nu}+\frac{p_1\cdot k}{p_2\cdot k}p_2^{\mu}p_2^{\nu}-p_1^{\mu}p_2^{\nu}-p_1^{\nu}p_2^{\mu}\bigg),\\
&\mathcal{B}_{(0)}^{\mu\nu}(p_f)=\kappa G_N\frac{\sigma(2\sigma^2-3)}{(\sigma^2-1)^{\frac{3}{2}}}\bigg(\frac{p_4\cdot k}{p_3\cdot k}p_3^{\mu}p_3^{\nu}+\frac{p_3\cdot k}{p_4\cdot k}p_4^{\mu}p_4^{\nu}-p_3^{\mu}p_4^{\nu}-p_3^{\nu}p_4^{\mu}\bigg), 
\end{align}
together with
\begin{align}
& \mathcal{B}_{(1)}^{\mu\nu}(p_i)=-i\kappa G_N\frac{\sigma(2\sigma^2-3)}{(\sigma^2-1)^{\frac{3}{2}}}\bigg[(b_1\cdot k)\frac{p_2\cdot k}{p_1\cdot k}p_1^{\mu}p_1^{\nu}+(b_2\cdot k)\frac{p_1\cdot k}{p_2\cdot k}p_2^{\mu}p_2^{\nu}\\&-(b_1\cdot k+b_2\cdot k)\frac{p_1^{\mu}p_2^{\nu}+p_1^{\nu}p_2^{\mu}}{2}-(p_2\cdot k)\frac{p_1^{\mu}b^{\nu}+p_1^{\nu}b^{\mu}
}{2}+(p_1\cdot k)\frac{p_2^{\mu}b^{\nu}+p_2^{\nu}b^{\mu}}{2}\bigg],
\end{align}
\begin{align}
\label{A6.1}
\nonumber&\mathcal{B}_{(1)}^{\mu\nu}(p_f)=-i\kappa G_N\frac{\sigma(2\sigma^2-3)}{(\sigma^2-1)^{\frac{3}{2}}}\bigg[(b_2\cdot k)\frac{p_4\cdot k}{p_3\cdot k}p_3^{\mu}p_3^{\nu}+(b_1\cdot k)\frac{p_4\cdot k}{p_3\cdot k}p_4^{\mu}p_4^{\nu}\\&-(b_1\cdot k+b_2\cdot k)\frac{p_3^{\mu}p_4^{\nu}+p_3^{\nu}p_4^{\nu}}{2}+(p_4\cdot k)\frac{p_3^{\mu}b^{\nu}+p_3^{\nu}b^{\mu}
}{2}+(p_3\cdot k)\frac{p_4^{\mu}b^{\nu}+p_4^{\nu}b^{\mu}}{2}\bigg],
\\\nonumber&\mathcal{C}^{\mu\nu}_{(1)}(p_i)=-\kappa G_N\frac{3(p_1\cdot k+p_2\cdot k)}{2m_1m_2(\sigma^2-1)^{\frac{5}{2}}}\bigg(\frac{p_2\cdot k}{p_1\cdot k}p_1^{\mu}p_1^{\nu}+\frac{p_1\cdot k}{p_2\cdot k}p_2^{\mu}p_2^{\nu}\\&-p_1^{\mu}p_2^{\nu}+p_1^{\nu}p_2^{\mu}\bigg),\\&\nonumber \mathcal{C}^{\mu\nu}_{(1)}(p_f)=-\kappa G_N\frac{3(p_3\cdot k+p_4\cdot k)}{2m_1m_2(\sigma^2-1)^{\frac{5}{2}}}\bigg(\frac{p_4\cdot k}{p_3\cdot k}p_3^{\mu}p_3^{\nu}+\frac{p_3\cdot k}{p_4\cdot k}p_4^{\mu}p_4^{\nu}\\&-p_3^{\mu}p_4^{\nu}+p_3^{\nu}p_4^{\mu}\bigg),
\end{align}
and with 
\begin{align}
\nonumber&\mathcal{D}_{(1)}^{\mu\nu}(p_i)=i\frac{\kappa}{2} G^2_N\bigg[\frac{\sigma(2\sigma^2-3)}{(\sigma^2-1)^{\frac{3}{2}}}\bigg]^2(p_1\cdot k+p_2\cdot k)\bigg(\frac{p_2\cdot k}{p_1\cdot k}p_1^{\mu}p_1^{\nu}+\frac{p_1\cdot k}{p_2\cdot k}p_2^{\mu}p_2^{\nu}\\&-p_1^{\mu}p_2^{\nu}-p_1^{\nu}p_2^{\mu}\bigg),\\&\nonumber \mathcal{D}_{(1)}^{\mu\nu}(p_f)=i\frac{\kappa}{2} G^2_N\bigg[\frac{\sigma(2\sigma^2-3)}{(\sigma^2-1)^{\frac{3}{2}}}\bigg]^2(p_3\cdot k+p_4\cdot k)\bigg(\frac{p_4\cdot k}{p_3\cdot k}p_3^{\mu}p_3^{\nu}+\frac{p_3\cdot k}{p_4\cdot k}p_4^{\mu}p_4^{\nu}\\&-p_3^{\mu}p_4^{\nu}-p_3^{\nu}p_4^{\mu}\bigg).
\end{align}
All the tensorial structures appearing above automatically satisfy gauge invariance $k_{\mu}\mathcal{T}^{\mu\nu}_{(a)}=k_{\nu}\mathcal{T}^{\mu\nu}_{(a)}=0$. We can PM expand the tensors that depend on the outgoing momenta $p_f$ as
\begin{align}
\label{A8}
\nonumber&\mathcal{B}_{(0)}^{\mu\nu}(p_f)=\mathcal{B}_{(0)}^{\mu\nu}(p_i)-2\kappa\frac{G^2_Nm_1m_2}{b^2}(p_1\cdot k+p_2\cdot k)\frac{\sigma(2\sigma^2-3)}{(\sigma^2-1)^{\frac{3}{2}}}\frac{2\sigma^2-1}{\sqrt{\sigma^2-1}}\\&\times \bigg[(b\cdot k)\bigg(\frac{p_1^{\mu}p_1^{\nu}}{(p_1\cdot k)^2}-\frac{p_2^{\mu}p_2^{\nu}}{(p_2\cdot k)^2}\bigg)-\frac{p_1^{\mu}b^{\nu}+p_1^{\nu}b^{\mu}}{p_1\cdot k}+\frac{p_2^{\mu}b^{\nu}+p_2^{\nu}b^{\mu}}{p_2\cdot k}\bigg]+\mathcal{O}(G^3_N),\\
\nonumber &\mathcal{B}_{(1)}^{\mu\nu}(p_f)=\mathcal{B}_{(1)}^{\mu\nu}(p_i)+2i\kappa \frac{G^2_N m_1m_2}{b^2}\frac{\sigma(2\sigma^2-3)}{(\sigma^2-1)^{\frac{3}{2}}}\frac{2\sigma^2-1}{\sqrt{\sigma^2-1}}(p_1\cdot k+p_2\cdot k)\bigg[\\\nonumber&(b\cdot k) \bigg((b_1\cdot k)\frac{p_1^{\mu}p_1^{\nu}}{(p_1\cdot k)^2}-(b_2\cdot k)\frac{p_2^{\mu}p_2^{\nu}}{(p_2\cdot k)^2}\bigg)-(b_1\cdot k)\frac{p_1^{\mu}b^{\nu}+p_1^{\nu}b^{\mu}}{p_1\cdot k}\\&\label{A8.1}+(b_2\cdot k)\frac{p_2^{\mu}b^{\nu}+p_2^{\nu}b^{\mu}}{p_2\cdot k}+b^{\mu}b^{\nu}\bigg]+\mathcal{O}(G^3_N),\\
&\nonumber \mathcal{C}_{(1)}^{\mu\nu}(p_f)=-\mathcal{C}_{(1)}^{\mu\nu}(p_i)-\kappa G^2_N\frac{3}{(\sigma^2-1)^{3}}\frac{(p_1\cdot k+p_2\cdot k)^2}{b^2}\bigg[(b\cdot k)\bigg(\frac{p_1^{\mu}p_1^{\nu}}{(p_1\cdot k)^2}-\frac{p_2^{\mu}p_2^{\nu}}{(p_2\cdot k)^2}\bigg)\\\label{A10}&-\bigg(\frac{p_1^{\mu}b^{\nu}+p_1^{\nu}b^{\mu}}{p_1\cdot k}-\frac{p_2^{\mu}b^{\nu}+p_2^{\nu}b^{\mu}}{p_2\cdot k}\bigg)\bigg]+\mathcal{O}(G^3_N),\\ &\mathcal{D}_{(1)}^{\mu\nu}(p_f)=- \mathcal{D}_{(1)}^{\mu\nu}(p_i)+\mathcal{O}(G^3_N).
\end{align}
\section{Classical limit of the infrared divergent factor}
\label{C}

In this appendix we extract the classical limit of the infrared divergent factor ${\cal{W}}$ in \eqref{W1}  that appears at one-loop level and that, once exponentiated, takes care of all infrared divergences at arbitrary perturbative order. It is a one-loop diagram involving two massive propagators and a graviton propagator that is attached to the two massive lines.

It was originally computed by Weinberg~\cite{Weinberg:1965nx} and it is given by
\begin{align}
{\cal{W}}=  \frac{G}{\pi}  \log \frac{\lambda}{\Lambda} \sum_{j,k=1}^N  m_j m_k \eta_j \eta_k\bigg (\sigma_{jk}^2 - \frac{1}{2}\bigg)\frac{ \log \left( \sigma_{jk} + \sqrt{\sigma^2_{jk} -1}\right) -i \pi 
\delta_{\eta_j \eta_k,1}  }{\sqrt{\sigma^2_{jk}-1}}
\label{C1}
\end{align}
where $\Lambda$    $(\lambda)$ is an ultraviolet (infrared) cutoff and the sums are extended over the total number $N$ of  particles participating in the scattering. For $j=k$ the internal massless line connects two points of the same massive line and therefore it does not contribute to the classical amplitude. In this case, the correct procedure~\cite{Yennie:1961ad} is to simply drop the divergent imaginary part and to compute the real part for $\sigma_{jj} \rightarrow 1$. For these terms of the sum with $j=k$ one gets just $\frac{m_j^2}{2}$.

Equation \eqref{C1} has also been  computed using dimensional regularisation (for instance in Ref.~\cite{Heissenberg:2021tzo}) and the two results match each other
with  the substitution
\begin{equation}
\log \frac{\lambda}{\Lambda} \longrightarrow \frac{1}{2\epsilon}
    \label{C2},
\end{equation}
 similarly to \eqref{conv}. By rewriting \eqref{C1} in the following form
\begin{align}
\nonumber
{\cal{W}}=&  \frac{G}{\pi}  \log \frac{\lambda}{\Lambda} \sum_{j,k=1}^N    \frac{(p_j \cdot p_k)^2 - \frac{1}{2}m_j^2 m_k^2}{\sqrt{(p_j\cdot p_k)^2-m_j^2 m_k^2}} \Bigg[ \frac{\eta_j \eta_k}{2}\log \frac{-\eta_j \eta_k p_j\cdot p_k + \sqrt{(p_j\cdot p_k)^2 -m_j^2 m_k^2}}{-\eta_j \eta_k p_j\cdot p_k - \sqrt{(p_j\cdot p_k)^2 -m_j^2 m_k^2}}\\& -i \pi 
\delta_{\eta_j \eta_k,1}  \Bigg],
\label{C3}
\end{align}
where the quantity $- \eta_j \eta_k p_j\cdot p_k$ is always positive, we can take the limit where the particle with $j,k=N$ is massless.
In this case we get the same expression as in \eqref{C3} where, however, the indices $j,k$ go from $1$ to $N-1$ and an additional term, containing the momentum $p_N=k$, given by~\cite{Georgoudis:2023eke}
\begin{equation}
{\cal{W}}_{\mathrm{phase}} = \frac{G}{\pi}  \log \frac{\lambda}{\Lambda}\sum_{j=1}^{N-1} 2(-p_j \cdot k) \bigg[ \frac{1}{2} \log \frac{ 4 (p_j\cdot k)^2}{m_j^2 \mu^2} - i \pi \delta_{\eta_j \eta_N,1} \bigg].
\label{C4}
\end{equation}
As explained in Ref.~\cite{Georgoudis:2023eke} the terms with $\log m_N$ 
cancel because of momentum conservation leaving behind an arbitrary energy scale $\mu$.

In the final part of this appendix we  show that the terms with the logarithms in \eqref{C3} and \eqref{C4} are negligible in the classical limit.

Let us consider for simplicity the case with $N=4$ with two incoming and two outgoing massive particles in \eqref{C3} with mass $m_1$ and $m_2$ respectively. We have four types of logarithmic terms in \eqref{C3}. Two of them, corresponding to $(jk)= (12)(21)(34)(43)$ and $(jk)=(13)(31)(24)(42)$,  can be respectively written in terms of the Mandelstam variables $s$ and $u$. Since, at $t\sim 0$, $s-m_1^2-m_2^2 =- (u-m_1^2 -m_2^2)$, one can see that their sum is proportional to $t$ and vanishes for $t=0$. Four terms, corresponding to $(jk)=(11)(22)(33)(44)$, give a contribution  to the sum equal to $m_1^2+m_2^2$. The remaining four terms, corresponding to $(jk)=(14)(41)(23)(32)$, can be written in terms of the Mandelstam variable $t$ and for small $t$ they give a contribution equal to $m_1^2+m_2^2$ plus terms of order $t$. In conclusion, the sum of the 16 terms behaves as $t$, for small $t$, that is of the order $\hbar^2$ 
and therefore negligible in the classical limit.    

In order to show that also the sum of logarithmic terms in \eqref{C4} are quantum one introduces the two quantities $Q_1 =p_1 +p_4$ and $Q_2=p_2+p_3$ and rewrites the sum in terms of $p_1, p_2, Q_1$ and $Q_2$
obtaining a quantity proportional to $Q_1$ and $Q_2$ for small $Q_1$ and $Q_2$ that is of the order $\hbar$ and therefore negligible in the classical limit.

Notice that, with the choice $\lambda=\omega$, \eqref{C3} is equal to Eq. (1.5) of \cite{Krishna:2023fxg} except for the sum where the terms $j = k$ are excluded and 
the second term of \eqref{C4} corresponds to the quantity in Eq. (1.6) of 
\cite{Krishna:2023fxg} apart from some signs probably due to different conventions.

\bibliographystyle{utphys}
\bibliography{hie4.bib}

\end{document}